\documentclass[a4paper,12pt]{article}
\usepackage{amsfonts}
\usepackage{amsmath}
\usepackage{amssymb}
\usepackage{graphicx}

\makeatletter
 
\def\diagram{\m@th\leftwidth=\z@ \rightwidth=\z@ \topheight=\z@
\botheight=\z@ \setbox\@picbox\hbox\bgroup}
 
\def\enddiagram{\egroup\wd\@picbox\rightwidth\unitlength
\ht\@picbox\topheight\unitlength \dp\@picbox\botheight\unitlength
\hskip\leftwidth\unitlength\box\@picbox}
 
\def\bfig{\begin{diagram}}
\def\efig{\end{diagram}}
\newcount\wideness \newcount\leftwidth \newcount\rightwidth
\newcount\highness \newcount\topheight \newcount\botheight
 
\def\ratchet#1#2{\ifnum#1<#2 \global #1=#2 \fi}
 
\def\putbox(#1,#2)#3{%
\horsize{\wideness}{#3} \divide\wideness by 2
{\advance\wideness by #1 \ratchet{\rightwidth}{\wideness}}
{\advance\wideness by -#1 \ratchet{\leftwidth}{\wideness}}
\vertsize{\highness}{#3} \divide\highness by 2
{\advance\highness by #2 \ratchet{\topheight}{\highness}}
{\advance\highness by -#2 \ratchet{\botheight}{\highness}}
\put(#1,#2){\makebox(0,0){$#3$}}}
 
\def\putlbox(#1,#2)#3{%
\horsize{\wideness}{#3}
{\advance\wideness by #1 \ratchet{\rightwidth}{\wideness}}
{\ratchet{\leftwidth}{-#1}}
\vertsize{\highness}{#3} \divide\highness by 2
{\advance\highness by #2 \ratchet{\topheight}{\highness}}
{\advance\highness by -#2 \ratchet{\botheight}{\highness}}
\put(#1,#2){\makebox(0,0)[l]{$#3$}}}
 
\def\putrbox(#1,#2)#3{%
\horsize{\wideness}{#3}
{\ratchet{\rightwidth}{#1}}
{\advance\wideness by -#1 \ratchet{\leftwidth}{\wideness}}
\vertsize{\highness}{#3} \divide\highness by 2
{\advance\highness by #2 \ratchet{\topheight}{\highness}}
{\advance\highness by -#2 \ratchet{\botheight}{\highness}}
\put(#1,#2){\makebox(0,0)[r]{$#3$}}}

\def\adjust[#1]{} 
 
\newcount \coefa
\newcount \coefb
\newcount \coefc
\newcount\tempcounta
\newcount\tempcountb
\newcount\tempcountc
\newcount\tempcountd
\newcount\xext
\newcount\yext
\newcount\xoff
\newcount\yoff
\newcount\gap%
\newcount\arrowtypea
\newcount\arrowtypeb
\newcount\arrowtypec
\newcount\arrowtyped
\newcount\arrowtypee
\newcount\height
\newcount\width
\newcount\xpos
\newcount\ypos
\newcount\run
\newcount\rise
\newcount\arrowlength
\newcount\halflength
\newcount\arrowtype
\newdimen\tempdimen
\newdimen\xlen
\newdimen\ylen
\newsavebox{\tempboxa}%
\newsavebox{\tempboxb}%
\newsavebox{\tempboxc}%
 
\newdimen\w@dth
 
\def\setw@dth#1#2{\setbox\z@\hbox{\m@th$#1$}\w@dth=\wd\z@
\setbox\@ne\hbox{\m@th$#2$}\ifnum\w@dth<\wd\@ne \w@dth=\wd\@ne \fi
\advance\w@dth by 1.2em}
 
\def\t@^#1_#2{\allowbreak\def\n@one{#1}\def\n@two{#2}\mathrel
{\setw@dth{#1}{#2}
\mathop{\hbox to \w@dth{\rightarrowfill}}\limits
\ifx\n@one\empty\else ^{\box\z@}\fi
\ifx\n@two\empty\else _{\box\@ne}\fi}}
\def\t@@^#1{\@ifnextchar_{\t@^{#1}}{\t@^{#1}_{}}}
\def\to{\@ifnextchar^{\t@@}{\t@@^{}}}
 
\def\t@left^#1_#2{\def\n@one{#1}\def\n@two{#2}\mathrel{\setw@dth{#1}{#2}
\mathop{\hbox to \w@dth{\leftarrowfill}}\limits
\ifx\n@one\empty\else ^{\box\z@}\fi
\ifx\n@two\empty\else _{\box\@ne}\fi}}
\def\t@@left^#1{\@ifnextchar_{\t@left^{#1}}{\t@left^{#1}_{}}}
\def\toleft{\@ifnextchar^{\t@@left}{\t@@left^{}}}
 
\def\two@^#1_#2{\allowbreak
\def\n@one{#1}\def\n@two{#2}\mathrel{\setw@dth{#1}{#2}
\mathop{\vcenter{\lineskip\z@\baselineskip\z@
                 \hbox to \w@dth{\rightarrowfill}%
                 \hbox to \w@dth{\rightarrowfill}}%
       }\limits
\ifx\n@one\empty\else ^{\box\z@}\fi
\ifx\n@two\empty\else _{\box\@ne}\fi}}
\def\tw@@^#1{\@ifnextchar _{\two@^{#1}}{\two@^{#1}_{}}}
\def\two{\@ifnextchar ^{\tw@@}{\tw@@^{}}}
 
\def\tofr@^#1_#2{\def\n@one{#1}\def\n@two{#2}\mathrel{\setw@dth{#1}{#2}
\mathop{\vcenter{\hbox to \w@dth{\rightarrowfill}\kern-1.7ex
                 \hbox to \w@dth{\leftarrowfill}}%
       }\limits
\ifx\n@one\empty\else ^{\box\z@}\fi
\ifx\n@two\empty\else _{\box\@ne}\fi}}
\def\t@fr@^#1{\@ifnextchar_ {\tofr@^{#1}}{\tofr@^{#1}_{}}}
\def\tofro{\@ifnextchar^ {\t@fr@}{\t@fr@^{}}}

\def\mon{\mathop{\m@th\hbox to
      14.6\P@{\lasyb\char'51\hskip-2.1\P@$\arrext$\hss
$\mathord\rightarrow$}}\limits} 
\def\leftmono{\mathrel{\m@th\hbox to
14.6\P@{$\mathord\leftarrow$\hss$\arrext$\hskip-2.1\P@\lasyb\char'50%
}}\limits} 
\mathchardef\arrext="0200       

\def\settypes(#1,#2,#3){\arrowtypea#1 \arrowtypeb#2 \arrowtypec#3}
\def\settoheight#1#2{\setbox\@tempboxa\hbox{#2}#1\ht\@tempboxa\relax}%
\def\settodepth#1#2{\setbox\@tempboxa\hbox{#2}#1\dp\@tempboxa\relax}%
\def\settokens`#1`#2`#3`#4`{%
     \def\tokena{#1}\def\tokenb{#2}\def\tokenc{#3}\def\tokend{#4}}
\def\setsqparms[#1`#2`#3`#4;#5`#6]{%
\arrowtypea #1
\arrowtypeb #2
\arrowtypec #3
\arrowtyped #4
\width #5
\height #6
}
\def\setpos(#1,#2){\xpos=#1 \ypos#2}

\def\settriparms[#1`#2`#3;#4]{\settripairparms[#1`#2`#3`1`1;#4]}%
 
\def\settripairparms[#1`#2`#3`#4`#5;#6]{%
\arrowtypea #1
\arrowtypeb #2
\arrowtypec #3
\arrowtyped #4
\arrowtypee #5
\width #6
\height #6
}
 
\def\resetparms{\settripairparms[1`1`1`1`1;500]\width 500}
 
\resetparms
 
\def\mvector(#1,#2)#3{
\put(0,0){\vector(#1,#2){#3}}%
\put(0,0){\vector(#1,#2){26}}%
}
\def\evector(#1,#2)#3{{
\arrowlength #3
\put(0,0){\vector(#1,#2){\arrowlength}}%
\advance \arrowlength by-30
\put(0,0){\vector(#1,#2){\arrowlength}}%
}}
 
\def\horsize#1#2{%
\settowidth{\tempdimen}{$#2$}%
#1=\tempdimen
\divide #1 by\unitlength
}
 
\def\vertsize#1#2{%
\settoheight{\tempdimen}{$#2$}%
#1=\tempdimen
\settodepth{\tempdimen}{$#2$}%
\advance #1 by\tempdimen
\divide #1 by\unitlength
}
 
\def\putvector(#1,#2)(#3,#4)#5#6{{%
\ifnum3<\arrowtype
\putdashvector(#1,#2)(#3,#4)#5\arrowtype
\else
\ifnum\arrowtype<-3
\putdashvector(#1,#2)(#3,#4)#5\arrowtype
\else
\xpos=#1
\ypos=#2
\run=#3
\rise=#4
\arrowlength=#5
\ifnum \arrowtype<0
    \ifnum \run=0
        \advance \ypos by-\arrowlength
    \else
        \tempcounta \arrowlength
        \multiply \tempcounta by\rise
        \divide \tempcounta by\run
        \ifnum\run>0
            \advance \xpos by\arrowlength
            \advance \ypos by\tempcounta
        \else
            \advance \xpos by-\arrowlength
            \advance \ypos by-\tempcounta
        \fi
    \fi
    \multiply \arrowtype by-1
    \multiply \rise by-1
    \multiply \run by-1
\fi
\ifcase \arrowtype
\or \put(\xpos,\ypos){\vector(\run,\rise){\arrowlength}}%
\or \put(\xpos,\ypos){\mvector(\run,\rise)\arrowlength}%
\or \put(\xpos,\ypos){\evector(\run,\rise){\arrowlength}}%
\fi\fi\fi
}}
 
\def\putsplitvector(#1,#2)#3#4{
\xpos #1
\ypos #2
\arrowtype #4
\halflength #3
\arrowlength #3
\gap 140
\advance \halflength by-\gap
\divide \halflength by2
\ifnum\arrowtype>0
   \ifcase \arrowtype
   \or \put(\xpos,\ypos){\line(0,-1){\halflength}}%
       \advance\ypos by-\halflength
       \advance\ypos by-\gap
       \put(\xpos,\ypos){\vector(0,-1){\halflength}}%
   \or \put(\xpos,\ypos){\line(0,-1)\halflength}%
       \put(\xpos,\ypos){\vector(0,-1)3}%
       \advance\ypos by-\halflength
       \advance\ypos by-\gap
       \put(\xpos,\ypos){\vector(0,-1){\halflength}}%
   \or \put(\xpos,\ypos){\line(0,-1)\halflength}%
       \advance\ypos by-\halflength
       \advance\ypos by-\gap
       \put(\xpos,\ypos){\evector(0,-1){\halflength}}%
   \fi
\else \arrowtype=-\arrowtype
   \ifcase\arrowtype
   \or \advance \ypos by-\arrowlength
       \put(\xpos,\ypos){\line(0,1){\halflength}}%
       \advance\ypos by\halflength
       \advance\ypos by\gap
       \put(\xpos,\ypos){\vector(0,1){\halflength}}%
   \or \advance \ypos by-\arrowlength
       \put(\xpos,\ypos){\line(0,1)\halflength}%
       \put(\xpos,\ypos){\vector(0,1)3}%
       \advance\ypos by\halflength
       \advance\ypos by\gap
       \put(\xpos,\ypos){\vector(0,1){\halflength}}%
   \or \advance \ypos by-\arrowlength
       \put(\xpos,\ypos){\line(0,1)\halflength}%
       \advance\ypos by\halflength
       \advance\ypos by\gap
       \put(\xpos,\ypos){\evector(0,1){\halflength}}%
   \fi
\fi
}
 
\def\putmorphism(#1)(#2,#3)[#4`#5`#6]#7#8#9{{%
\run #2
\rise #3
\ifnum\rise=0
  \puthmorphism(#1)[#4`#5`#6]{#7}{#8}#9%
\else\ifnum\run=0
  \putvmorphism(#1)[#4`#5`#6]{#7}{#8}#9%
\else
\setpos(#1)%
\arrowlength #7
\arrowtype #8
\ifnum\run=0
\else\ifnum\rise=0
\else
\ifnum\run>0
    \coefa=1
\else
   \coefa=-1
\fi
\ifnum\arrowtype>0
   \coefb=0
   \coefc=-1
\else
   \coefb=\coefa
   \coefc=1
   \arrowtype=-\arrowtype
\fi
\width=2
\multiply \width by\run
\divide \width by\rise
\ifnum \width<0  \width=-\width\fi
\advance\width by60
\if l#9 \width=-\width\fi
\putbox(\xpos,\ypos){#4}
{\multiply \coefa by\arrowlength
\advance\xpos by\coefa
\multiply \coefa by\rise
\divide \coefa by\run
\advance \ypos by\coefa
\putbox(\xpos,\ypos){#5} }%
{\multiply \coefa by\arrowlength
\divide \coefa by2
\advance \xpos by\coefa
\advance \xpos by\width
\multiply \coefa by\rise
\divide \coefa by\run
\advance \ypos by\coefa
\if l#9%
   \putrbox(\xpos,\ypos){#6}%
\else\if r#9%
   \putlbox(\xpos,\ypos){#6}%
\fi\fi }%
{\multiply \rise by-\coefc
\multiply \run by-\coefc
\multiply \coefb by\arrowlength
\advance \xpos by\coefb
\multiply \coefb by\rise
\divide \coefb by\run
\advance \ypos by\coefb
\multiply \coefc by70
\advance \ypos by\coefc
\multiply \coefc by\run
\divide \coefc by\rise
\advance \xpos by\coefc
\multiply \coefa by140
\multiply \coefa by\run
\divide \coefa by\rise
\advance \arrowlength by\coefa
\ifcase\arrowtype
\or \put(\xpos,\ypos){\vector(\run,\rise){\arrowlength}}%
\or \put(\xpos,\ypos){\mvector(\run,\rise){\arrowlength}}%
\or \put(\xpos,\ypos){\evector(\run,\rise){\arrowlength}}%
\fi}\fi\fi\fi\fi}}

\newcount\numbdashes \newcount\lengthdash \newcount\increment
 
\def\howmanydashes{
\numbdashes=\arrowlength \lengthdash=40
\divide\numbdashes by \lengthdash
\lengthdash=\arrowlength
\divide\lengthdash by \numbdashes
\increment=\lengthdash
\multiply\lengthdash by 3
\divide\lengthdash by 5
}
 
\def\putdashvector(#1)(#2,#3)#4#5{%
\ifnum#3=0 \putdashhvector(#1){#4}#5
\else
\ifnum#2=0
\putdashvvector(#1){#4}#5\fi\fi}
 
\def\putdashhvector(#1,#2)#3#4{{%
\arrowlength=#3 \howmanydashes
\multiput(#1,#2)(\increment,0){\numbdashes}%
{\vrule height .4pt width \lengthdash\unitlength}
\arrowtype=#4 \xpos=#1
\ifnum\arrowtype<0 \advance\arrowtype by 7 \fi
\ifcase\arrowtype
\or \advance\xpos by 10
    \put(\xpos,#2){\vector(-1,0){\lengthdash}}
    \advance\xpos by 40
    \put(\xpos,#2){\vector(-1,0){\lengthdash}}
\or \advance \xpos by 10
    \put(\xpos,#2){\vector(-1,0){\lengthdash}}
    \advance\xpos by  \arrowlength
    \advance\xpos by  -50
    \put(\xpos,#2){\vector(-1,0){\lengthdash}}
\or \advance\xpos by 10
    \put(\xpos,#2){\vector(-1,0){\lengthdash}}
\or \advance\xpos by \arrowlength
    \advance\xpos by -\lengthdash
    \put(\xpos,#2){\vector(1,0){\lengthdash}}
\or {\advance\xpos by 10
    \put(\xpos,#2){\vector(1,0){\lengthdash}}}
    \advance\xpos by \arrowlength
    \advance\xpos by -\lengthdash
    \put(\xpos,#2){\vector(1,0){\lengthdash}}
\or \advance\xpos by \arrowlength
    \advance\xpos by -\lengthdash
    \put(\xpos,#2){\vector(1,0){\lengthdash}}
    \advance\xpos by -40
    \put(\xpos,#2){\vector(1,0){\lengthdash}}
   \fi
}}
 
\def\putdashvvector(#1,#2)#3#4{{%
\arrowlength=#3 \howmanydashes
\ypos=#2 \advance\ypos by -\arrowlength
\multiput(#1,#2)(0,\increment){\numbdashes}%
    {\vrule width .4pt height \lengthdash\unitlength}
\arrowtype=#4 \ypos=#2
\ifnum\arrowtype<0 \advance\arrowtype by 7 \fi
\ifcase\arrowtype
\or \advance\ypos by \arrowlength \advance\ypos by -40
    \put(#1,\ypos){\vector(0,1){\lengthdash}}
    \advance\ypos by -40
    \put(#1,\ypos){\vector(0,1){\lengthdash}}
\or \advance\ypos by 10
    \put(#1,\ypos){\vector(0,1){\lengthdash}}
    \advance\ypos by \arrowlength \advance\ypos by -40
    \put(#1,\ypos){\vector(0,1){\lengthdash}}
\or \advance\ypos by \arrowlength \advance\ypos by -40
    \put(#1,\ypos){\vector(0,1){\lengthdash}}
\or \advance\ypos by 10
    \put(#1,\ypos){\vector(0,-1){\lengthdash}}
\or \advance\ypos by 10
    \put(#1,\ypos){\vector(0,-1){\lengthdash}}
    \advance\ypos by \arrowlength \advance\ypos by -40
    \put(#1,\ypos){\vector(0,-1){\lengthdash}}
\or \advance\ypos by 10
    \put(#1,\ypos){\vector(0,-1){\lengthdash}}
    \advance\ypos by 40
    \put(#1,\ypos){\vector(0,-1){\lengthdash}}
\fi
}}
 
\def\puthmorphism(#1,#2)[#3`#4`#5]#6#7#8{{%
\xpos #1
\ypos #2
\width #6
\arrowlength #6
\arrowtype=#7
\putbox(\xpos,\ypos){#3\vphantom{#4}}%
{\advance \xpos by\arrowlength
\putbox(\xpos,\ypos){\vphantom{#3}#4}}%
\horsize{\tempcounta}{#3}%
\horsize{\tempcountb}{#4}%
\divide \tempcounta by2
\divide \tempcountb by2
\advance \tempcounta by30
\advance \tempcountb by30
\advance \xpos by\tempcounta
\advance \arrowlength by-\tempcounta
\advance \arrowlength by-\tempcountb
\putvector(\xpos,\ypos)(1,0)\arrowlength\arrowtype
\divide \arrowlength by2
\advance \xpos by\arrowlength
\vertsize{\tempcounta}{#5}%
\divide\tempcounta by2
\advance \tempcounta by20
\if a#8 %
   \advance \ypos by\tempcounta
   \putbox(\xpos,\ypos){#5}%
\else
   \advance \ypos by-\tempcounta
   \putbox(\xpos,\ypos){#5}%
\fi}}
 
\def\putvmorphism(#1,#2)[#3`#4`#5]#6#7#8{{%
\xpos #1
\ypos #2
\arrowlength #6
\arrowtype #7
\settowidth{\xlen}{$#5$}%
\putbox(\xpos,\ypos){#3}%
{\advance \ypos by-\arrowlength
\putbox(\xpos,\ypos){#4}}%
{\advance\arrowlength by-140
\advance \ypos by-70
\ifdim\xlen>0pt
   \if m#8%
      \putsplitvector(\xpos,\ypos)\arrowlength\arrowtype
   \else
   \putvector(\xpos,\ypos)(0,-1)\arrowlength\arrowtype
   \fi
\else
   \putvector(\xpos,\ypos)(0,-1)\arrowlength\arrowtype
\fi}%
\ifdim\xlen>0pt
   \divide \arrowlength by2
   \advance\ypos by-\arrowlength
   \if l#8%
      \advance \xpos by-40
      \putrbox(\xpos,\ypos){#5}%
   \else\if r#8%
      \advance \xpos by40
      \putlbox(\xpos,\ypos){#5}%
   \else
      \putbox(\xpos,\ypos){#5}%
   \fi\fi
\fi
}}
 
\def\putsquarep<#1>(#2)[#3;#4`#5`#6`#7]{{%
\setsqparms[#1]%
\setpos(#2)%
\settokens`#3`%
\puthmorphism(\xpos,\ypos)[\tokenc`\tokend`{#7}]{\width}{\arrowtyped}b%
\advance\ypos by \height
\puthmorphism(\xpos,\ypos)[\tokena`\tokenb`{#4}]{\width}{\arrowtypea}a%
\putvmorphism(\xpos,\ypos)[``{#5}]{\height}{\arrowtypeb}l%
\advance\xpos by \width
\putvmorphism(\xpos,\ypos)[``{#6}]{\height}{\arrowtypec}r%
}}
 
\def\putsquare{\@ifnextchar <{\putsquarep}{\putsquarep%
   <\arrowtypea`\arrowtypeb`\arrowtypec`\arrowtyped;\width`\height>}}
\def\square{\@ifnextchar< {\squarep}{\squarep
   <\arrowtypea`\arrowtypeb`\arrowtypec`\arrowtyped;\width`\height>}}
\def\squarep<#1>[#2`#3`#4`#5;#6`#7`#8`#9]{{
\setsqparms[#1]
\diagram
\putsquarep<\arrowtypea`\arrowtypeb`\arrowtypec`
\arrowtyped;\width`\height>
(0,0)[#2`#3`#4`{#5};#6`#7`#8`{#9}]
\enddiagram
}}                                                 
\def\putptrianglep<#1>(#2,#3)[#4`#5`#6;#7`#8`#9]{{%
\settriparms[#1]%
\xpos=#2 \ypos=#3
\advance\ypos by \height
\puthmorphism(\xpos,\ypos)[#4`#5`{#7}]{\height}{\arrowtypea}a%
\putvmorphism(\xpos,\ypos)[`#6`{#8}]{\height}{\arrowtypeb}l%
\advance\xpos by\height
\putmorphism(\xpos,\ypos)(-1,-1)[``{#9}]{\height}{\arrowtypec}r%
}}
 
\def\putptriangle{\@ifnextchar <{\putptrianglep}{\putptrianglep
   <\arrowtypea`\arrowtypeb`\arrowtypec;\height>}}
\def\ptriangle{\@ifnextchar <{\ptrianglep}{\ptrianglep
   <\arrowtypea`\arrowtypeb`\arrowtypec;\height>}}
\def\ptrianglep<#1>[#2`#3`#4;#5`#6`#7]{{
\settriparms[#1]
\diagram
\putptrianglep<\arrowtypea`\arrowtypeb`
\arrowtypec;\height>
(0,0)[#2`#3`#4;#5`#6`{#7}]
\enddiagram
}}                                            
 
\def\putqtrianglep<#1>(#2,#3)[#4`#5`#6;#7`#8`#9]{{%
\settriparms[#1]%
\xpos=#2 \ypos=#3
\advance\ypos by\height
\puthmorphism(\xpos,\ypos)[#4`#5`{#7}]{\height}{\arrowtypea}a%
\putmorphism(\xpos,\ypos)(1,-1)[``{#8}]{\height}{\arrowtypeb}l%
\advance\xpos by\height
\putvmorphism(\xpos,\ypos)[`#6`{#9}]{\height}{\arrowtypec}r%
}}
 
\def\putqtriangle{\@ifnextchar <{\putqtrianglep}{\putqtrianglep
   <\arrowtypea`\arrowtypeb`\arrowtypec;\height>}}
\def\qtriangle{\@ifnextchar <{\qtrianglep}{\qtrianglep
   <\arrowtypea`\arrowtypeb`\arrowtypec;\height>}}
\def\qtrianglep<#1>[#2`#3`#4;#5`#6`#7]{{
\settriparms[#1]
\width=\height                                
\diagram
\putqtrianglep<\arrowtypea`\arrowtypeb`
\arrowtypec;\height>
(0,0)[#2`#3`#4;#5`#6`{#7}]
\enddiagram
}}
 
\def\putdtrianglep<#1>(#2,#3)[#4`#5`#6;#7`#8`#9]{{%
\settriparms[#1]%
\xpos=#2 \ypos=#3
\puthmorphism(\xpos,\ypos)[#5`#6`{#9}]{\height}{\arrowtypec}b%
\advance\xpos by \height \advance\ypos by\height
\putmorphism(\xpos,\ypos)(-1,-1)[``{#7}]{\height}{\arrowtypea}l%
\putvmorphism(\xpos,\ypos)[#4``{#8}]{\height}{\arrowtypeb}r%
}}
 
\def\putdtriangle{\@ifnextchar <{\putdtrianglep}{\putdtrianglep
   <\arrowtypea`\arrowtypeb`\arrowtypec;\height>}}
\def\dtriangle{\@ifnextchar <{\dtrianglep}{\dtrianglep
   <\arrowtypea`\arrowtypeb`\arrowtypec;\height>}}
\def\dtrianglep<#1>[#2`#3`#4;#5`#6`#7]{{
\settriparms[#1]
\width=\height                                
\diagram
\putdtrianglep<\arrowtypea`\arrowtypeb`
\arrowtypec;\height>
(0,0)[#2`#3`#4;#5`#6`{#7}]
\enddiagram
}}
 
\def\putbtrianglep<#1>(#2,#3)[#4`#5`#6;#7`#8`#9]{{%
\settriparms[#1]%
\xpos=#2 \ypos=#3
\puthmorphism(\xpos,\ypos)[#5`#6`{#9}]{\height}{\arrowtypec}b%
\advance\ypos by\height
\putmorphism(\xpos,\ypos)(1,-1)[``{#8}]{\height}{\arrowtypeb}r%
\putvmorphism(\xpos,\ypos)[#4``{#7}]{\height}{\arrowtypea}l%
}}
 
\def\putbtriangle{\@ifnextchar <{\putbtrianglep}{\putbtrianglep
   <\arrowtypea`\arrowtypeb`\arrowtypec;\height>}}
\def\btriangle{\@ifnextchar <{\btrianglep}{\btrianglep
   <\arrowtypea`\arrowtypeb`\arrowtypec;\height>}}
\def\btrianglep<#1>[#2`#3`#4;#5`#6`#7]{{
\settriparms[#1]
\width=\height                               
\diagram
\putbtrianglep<\arrowtypea`\arrowtypeb`
\arrowtypec;\height>
(0,0)[#2`#3`#4;#5`#6`{#7}]
\enddiagram
}}
 
\def\putAtrianglep<#1>(#2,#3)[#4`#5`#6;#7`#8`#9]{{%
\settriparms[#1]%
\xpos=#2 \ypos=#3
{\multiply \height by2
\puthmorphism(\xpos,\ypos)[#5`#6`{#9}]{\height}{\arrowtypec}b}%
\advance\xpos by\height \advance\ypos by\height
\putmorphism(\xpos,\ypos)(-1,-1)[#4``{#7}]{\height}{\arrowtypea}l%
\putmorphism(\xpos,\ypos)(1,-1)[``{#8}]{\height}{\arrowtypeb}r%
}}
 
\def\putAtriangle{\@ifnextchar <{\putAtrianglep}{\putAtrianglep
   <\arrowtypea`\arrowtypeb`\arrowtypec;\height>}}
\def\Atriangle{\@ifnextchar <{\Atrianglep}{\Atrianglep
   <\arrowtypea`\arrowtypeb`\arrowtypec;\height>}}
\def\Atrianglep<#1>[#2`#3`#4;#5`#6`#7]{{
\settriparms[#1]
\width=\height                                     
\diagram
\putAtrianglep<\arrowtypea`\arrowtypeb`
\arrowtypec;\height>
(0,0)[#2`#3`#4;#5`#6`{#7}]
\enddiagram
}}
 
\def\putAtrianglepairp<#1>(#2)[#3;#4`#5`#6`#7`#8]{{%
\settripairparms[#1]%
\setpos(#2)%
\settokens`#3`%
\puthmorphism(\xpos,\ypos)[\tokenb`\tokenc`{#7}]{\height}{\arrowtyped}b%
\advance\xpos by\height
\puthmorphism(\xpos,\ypos)[\phantom{\tokenc}`\tokend`{#8}]%
{\height}{\arrowtypee}b%
\advance\ypos by\height
\putmorphism(\xpos,\ypos)(-1,-1)[\tokena``{#4}]{\height}{\arrowtypea}l%
\putvmorphism(\xpos,\ypos)[``{#5}]{\height}{\arrowtypeb}m%
\putmorphism(\xpos,\ypos)(1,-1)[``{#6}]{\height}{\arrowtypec}r%
}}
 
\def\putAtrianglepair{\@ifnextchar <{\putAtrianglepairp}{\putAtrianglepairp%
   <\arrowtypea`\arrowtypeb`\arrowtypec`\arrowtyped`\arrowtypee;\height>}}
\def\Atrianglepair{\@ifnextchar <{\Atrianglepairp}{\Atrianglepairp%
   <\arrowtypea`\arrowtypeb`\arrowtypec`\arrowtyped`\arrowtypee;\height>}}
 
\def\Atrianglepairp<#1>[#2;#3`#4`#5`#6`#7]{{
\settripairparms[#1]
\settokens`#2`
\width=\height                                
\diagram
\putAtrianglepairp                            
<\arrowtypea`\arrowtypeb`\arrowtypec`
\arrowtyped`\arrowtypee;\height>
(0,0)[{#2};#3`#4`#5`#6`{#7}]
\enddiagram
}}
 
\def\putVtrianglep<#1>(#2,#3)[#4`#5`#6;#7`#8`#9]{{%
\settriparms[#1]%
\xpos=#2 \ypos=#3
\advance\ypos by\height
{\multiply\height by2
\puthmorphism(\xpos,\ypos)[#4`#5`{#7}]{\height}{\arrowtypea}a}%
\putmorphism(\xpos,\ypos)(1,-1)[`#6`{#8}]{\height}{\arrowtypeb}l%
\advance\xpos by\height
\advance\xpos by\height
\putmorphism(\xpos,\ypos)(-1,-1)[``{#9}]{\height}{\arrowtypec}r%
}}
 
\def\putVtriangle{\@ifnextchar <{\putVtrianglep}{\putVtrianglep
   <\arrowtypea`\arrowtypeb`\arrowtypec;\height>}}
\def\Vtriangle{\@ifnextchar <{\Vtrianglep}{\Vtrianglep
   <\arrowtypea`\arrowtypeb`\arrowtypec;\height>}}
\def\Vtrianglep<#1>[#2`#3`#4;#5`#6`#7]{{
\settriparms[#1]
\width=\height                                 
\diagram
\putVtrianglep<\arrowtypea`\arrowtypeb`
\arrowtypec;\height>
(0,0)[#2`#3`#4;#5`#6`{#7}]
\enddiagram
}}
 
\def\putVtrianglepairp<#1>(#2)[#3;#4`#5`#6`#7`#8]{{
\settripairparms[#1]%
\setpos(#2)%
\settokens`#3`%
\advance\ypos by\height
\putmorphism(\xpos,\ypos)(1,-1)[`\tokend`{#6}]{\height}{\arrowtypec}l%
\puthmorphism(\xpos,\ypos)[\tokena`\tokenb`{#4}]{\height}{\arrowtypea}a%
\advance\xpos by\height
\puthmorphism(\xpos,\ypos)[\phantom{\tokenb}`\tokenc`{#5}]%
{\height}{\arrowtypeb}a%
\putvmorphism(\xpos,\ypos)[``{#7}]{\height}{\arrowtyped}m%
\advance\xpos by\height
\putmorphism(\xpos,\ypos)(-1,-1)[``{#8}]{\height}{\arrowtypee}r%
}}
 
\def\putVtrianglepair{\@ifnextchar <{\putVtrianglepairp}{\putVtrianglepairp%
    <\arrowtypea`\arrowtypeb`\arrowtypec`\arrowtyped`\arrowtypee;\height>}}
\def\Vtrianglepair{\@ifnextchar <{\Vtrianglepairp}{\Vtrianglepairp%
    <\arrowtypea`\arrowtypeb`\arrowtypec`\arrowtyped`\arrowtypee;\height>}}
\def\Vtrianglepairp<#1>[#2;#3`#4`#5`#6`#7]{{
\settripairparms[#1]
\settokens`#2`
\diagram
\putVtrianglepairp                             
<\arrowtypea`\arrowtypeb`\arrowtypec`
\arrowtyped`\arrowtypee;\height>
(0,0)[{#2};#3`#4`#5`#6`{#7}]
\enddiagram
}}

\def\putCtrianglep<#1>(#2,#3)[#4`#5`#6;#7`#8`#9]{{%
\settriparms[#1]%
\xpos=#2 \ypos=#3
\advance\ypos by\height
\putmorphism(\xpos,\ypos)(1,-1)[``{#9}]{\height}{\arrowtypec}l%
\advance\xpos by\height
\advance\ypos by\height
\putmorphism(\xpos,\ypos)(-1,-1)[#4`#5`{#7}]{\height}{\arrowtypea}l%
{\multiply\height by 2
\putvmorphism(\xpos,\ypos)[`#6`{#8}]{\height}{\arrowtypeb}r}%
}}
 
\def\putCtriangle{\@ifnextchar <{\putCtrianglep}{\putCtrianglep
    <\arrowtypea`\arrowtypeb`\arrowtypec;\height>}}
\def\Ctriangle{\@ifnextchar <{\Ctrianglep}{\Ctrianglep
    <\arrowtypea`\arrowtypeb`\arrowtypec;\height>}}
\def\Ctrianglep<#1>[#2`#3`#4;#5`#6`#7]{{
\settriparms[#1]
\width=\height                               
\diagram
\putCtrianglep<\arrowtypea`\arrowtypeb`
\arrowtypec;\height>
(0,0)[#2`#3`#4;#5`#6`{#7}]
\enddiagram
}}                                           
\def\putDtrianglep<#1>(#2,#3)[#4`#5`#6;#7`#8`#9]{{%
\settriparms[#1]%
\xpos=#2 \ypos=#3
\advance\xpos by\height \advance\ypos by\height
\putmorphism(\xpos,\ypos)(-1,-1)[``{#9}]{\height}{\arrowtypec}r%
\advance\xpos by-\height \advance\ypos by\height
\putmorphism(\xpos,\ypos)(1,-1)[`#5`{#8}]{\height}{\arrowtypeb}r%
{\multiply\height by 2
\putvmorphism(\xpos,\ypos)[#4`#6`{#7}]{\height}{\arrowtypea}l}%
}}
 
\def\putDtriangle{\@ifnextchar <{\putDtrianglep}{\putDtrianglep
    <\arrowtypea`\arrowtypeb`\arrowtypec;\height>}}
\def\Dtriangle{\@ifnextchar <{\Dtrianglep}{\Dtrianglep
   <\arrowtypea`\arrowtypeb`\arrowtypec;\height>}}
\def\Dtrianglep<#1>[#2`#3`#4;#5`#6`#7]{{
\settriparms[#1]
\width=\height                              
\diagram
\putDtrianglep<\arrowtypea`\arrowtypeb`
\arrowtypec;\height>
(0,0)[#2`#3`#4;#5`#6`{#7}]
\enddiagram
}}                                          
\def\setrecparms[#1`#2]{\width=#1 \height=#2}%
 
\def\recursep<#1`#2>[#3;#4`#5`#6`#7`#8]{{\m@th
\width=#1 \height=#2
\settokens`#3`
\settowidth{\tempdimen}{$\tokena$}
\ifdim\tempdimen=0pt
  \savebox{\tempboxa}{\hbox{$\tokenb$}}%
  \savebox{\tempboxb}{\hbox{$\tokend$}}%
  \savebox{\tempboxc}{\hbox{$#6$}}%
\else
  \savebox{\tempboxa}{\hbox{$\hbox{$\tokena$}\times\hbox{$\tokenb$}$}}%
  \savebox{\tempboxb}{\hbox{$\hbox{$\tokena$}\times\hbox{$\tokend$}$}}%
  \savebox{\tempboxc}{\hbox{$\hbox{$\tokena$}\times\hbox{$#6$}$}}%
\fi
\ypos=\height
\divide\ypos by 2
\xpos=\ypos
\advance\xpos by \width
\bfig
\putCtrianglep<-1`1`1;\ypos>(0,0)[`\tokenc`;#5`#6`{#7}]%
\puthmorphism(\ypos,0)[\tokend`\usebox{\tempboxb}`{#8}]{\width}{-1}b%
\puthmorphism(\ypos,\height)[\tokenb`\usebox{\tempboxa}`{#4}]{\width}{-1}a%
\advance\ypos by \width
\putvmorphism(\ypos,\height)[``\usebox{\tempboxc}]{\height}1r%
\efig
}}
 
\def\recurse{\@ifnextchar <{\recursep}{\recursep<\width`\height>}}
 
\def\puttwohmorphisms(#1,#2)[#3`#4;#5`#6]#7#8#9{{%
\puthmorphism(#1,#2)[#3`#4`]{#7}0a
\ypos=#2
\advance\ypos by 20
\puthmorphism(#1,\ypos)[\phantom{#3}`\phantom{#4}`#5]{#7}{#8}a
\advance\ypos by -40
\puthmorphism(#1,\ypos)[\phantom{#3}`\phantom{#4}`#6]{#7}{#9}b
}}
 
\def\puttwovmorphisms(#1,#2)[#3`#4;#5`#6]#7#8#9{{%
\putvmorphism(#1,#2)[#3`#4`]{#7}0a
\xpos=#1
\advance\xpos by -20
\putvmorphism(\xpos,#2)[\phantom{#3}`\phantom{#4}`#5]{#7}{#8}l
\advance\xpos by 40
\putvmorphism(\xpos,#2)[\phantom{#3}`\phantom{#4}`#6]{#7}{#9}r
}}
 
\def\puthcoequalizer(#1)[#2`#3`#4;#5`#6`#7]#8#9{{%
\setpos(#1)%
\puttwohmorphisms(\xpos,\ypos)[#2`#3;#5`#6]{#8}11%
\advance\xpos by #8
\puthmorphism(\xpos,\ypos)[\phantom{#3}`#4`#7]{#8}1{#9}
}}
 
\def\putvcoequalizer(#1)[#2`#3`#4;#5`#6`#7]#8#9{{%
\setpos(#1)%
\puttwovmorphisms(\xpos,\ypos)[#2`#3;#5`#6]{#8}11%
\advance\ypos by -#8
\putvmorphism(\xpos,\ypos)[\phantom{#3}`#4`#7]{#8}1{#9}
}}
 
\def\putthreehmorphisms(#1)[#2`#3;#4`#5`#6]#7(#8)#9{{%
\setpos(#1) \settypes(#8)
\if a#9 %
     \vertsize{\tempcounta}{#5}%
     \vertsize{\tempcountb}{#6}%
     \ifnum \tempcounta<\tempcountb \tempcounta=\tempcountb \fi
\else
     \vertsize{\tempcounta}{#4}%
     \vertsize{\tempcountb}{#5}%
     \ifnum \tempcounta<\tempcountb \tempcounta=\tempcountb \fi
\fi
\advance \tempcounta by 60
\puthmorphism(\xpos,\ypos)[#2`#3`#5]{#7}{\arrowtypeb}{#9}
\advance\ypos by \tempcounta
\puthmorphism(\xpos,\ypos)[\phantom{#2}`\phantom{#3}`#4]{#7}{\arrowtypea}{#9}
\advance\ypos by -\tempcounta \advance\ypos by -\tempcounta
\puthmorphism(\xpos,\ypos)[\phantom{#2}`\phantom{#3}`#6]{#7}{\arrowtypec}{#9}
}}
 
\def\setarrowtoks[#1`#2`#3`#4`#5`#6]{%
\def\toka{#1}
\def\tokb{#2}
\def\tokc{#3}
\def\tokd{#4}
\def\toke{#5}
\def\tokf{#6}
}
\def\hex{\@ifnextchar <{\hexp}{\hexp<1000`400>}}
\def\hexp<#1`#2>[#3`#4`#5`#6`#7`#8;#9]{%
\setarrowtoks[#9]
\yext=#2 \advance \yext by #2
\xext=#1 \advance\xext by \yext
\bfig
\putCtriangle<-1`0`1;#2>(0,0)[`#5`;\tokb``\tokd]
\xext=#1 \yext=#2 \advance \yext by #2
\putsquare<1`0`0`1;\xext`\yext>(#2,0)[#3`#4`#7`#8;\toka```\tokf]
\advance \xext by #2
\putDtriangle<0`1`-1;#2>(\xext,0)[`#6`;`\tokc`\toke]
\efig
}
\makeatother

\let\ssection=\section
\renewcommand{\section}{\setcounter{equation}{0}\ssection}

\newtheorem{definition}{Definition}[section]
\newtheorem{theorem}{Theorem}[section]
\newtheorem{lemma}[theorem]{Lemma}

\newenvironment{proof}[1][Proof]{\noindent\textbf{#1.} }{\ \rule{0.5em}{0.5em}}

\newcommand\mathC{\mkern1mu\raise2.2pt\hbox{$\scriptscriptstyle|$}
        {\mkern-7mu\rm C}}                
\newcommand{\mathR}{{\rm I\! R}}      

\newcommand\ie{{i.e.},}
\newcommand{\di}{\diamond}
\newcommand{\ga}{\gamma}
\newcommand{\Ga}{\Gamma}
\newcommand{\ka}{\kappa}
\renewcommand\l{\lambda}
\newcommand\s{\sigma}
\newcommand\Si{\Sigma}
\newcommand\De{\Delta}

\renewcommand{\O}{\Omega}

\newcommand{\id}{{\rm id}}
\newcommand\la{\langle}
\newcommand{\map}{\rightarrow}
\newcommand\ra{\rangle}
\newcommand{\sa}{{\rm sa}}
\newcommand\sq{\rightsquigarrow}
\newcommand{\op}{{\rm op}}

\newcommand\pr{{\rm pr}}                                                        

\newcommand{\A}{{\hat A}}

\renewcommand{\P}{{\hat P}}
\renewcommand{\S}{{\cal S}}
\newcommand{\Hi}{{\cal H}}

\newcommand\BH{\mathcal{B(H)}}
\newcommand\PH{\mathcal{P(H)}}

\newcommand\TO{\mathbb{T}}   

\newcommand\LeftDB{[\mkern-3mu[}
\newcommand\RightDB{]\mkern-3mu]}

\newcommand\eq[1]{(\ref{#1})}
\newcommand\eqs[2]{(\ref{#1}--\ref{#2})}
\newcommand\SAin[1]{\mbox{``}A\,\varepsilon\,#1\mbox{''}}

\newcommand\va[1]{\tilde{#1}}

\newcommand\dasto[2]{\delta(\hat{#2})_{#1}}     
\newcommand\dastoi[2]{\delta^i(\hat{#2})_{#1}}  

\newcommand\dasB[1]{\breve{\delta}(#1)}

\newcommand\dasBo[1]{\breve{\delta}^o(#1)}

\renewcommand\L[1]{\mathcal{L}({#1})}
\newcommand\PL[1]{{\cal PL}(#1)}
\renewcommand\sp[1]{{\rm sp}(\hat A)}

\newcommand\Val[1]{\LeftDB\,#1\,\RightDB}
\newcommand\TValM[1]{\nu(\,#1\,)}               
\newcommand\Hom[3]{{\rm Hom}_{#1}\big(#2,#3\big)}

\newcommand\name[1]{\ulcorner #1\urcorner}                

\newcommand\ps[1]{\underline{#1}}
\newcommand{\dG}{\ps{\mkern1mu\raise2.5pt\hbox{$\scriptscriptstyle|$}
        {\mkern-7mu\rm O}}}                 
\newcommand{\dOU}{\ps{\mkern1mu\raise2.5pt\hbox{$\scriptscriptstyle|$}
        {\mkern-7mu\rm U}}}               

\newcommand{\Sig}{\ps{\Sigma}}            
\newcommand{\R}{{\cal R}}                 

\newcommand{\SR}{\ps{{\mathR}^\succeq}}         

\newcommand\F[1]{F_{\L{#1}}\big(\Sigma,\R\big)}
\newcommand\Ob[1]{{\rm Ob(#1)}}
\newcommand\Sub[1]{{\rm Sub}(#1)}              

\newcommand{\Loc}{\bf Loc}
\newcommand\Set{{\bf Sets}}                    
\newcommand\SetH[1]{\Set^{{\V{#1}}^{\rm op}}}  
\newcommand\SetC[1]{\Set^{{#1}^{\rm op}}}      

\newcommand{\Sys}{{\bf Sys}}                   
\newcommand\V[1]{{\cal V}(\Hi_{#1})}           

\begin{document}

\begin{titlepage}

\begin{center}
{\large\bf A Topos Foundation for Theories of Physics:\\[6pt]
IV. Categories of Systems}
\end{center}

\vspace{0.8 truecm}
\begin{center}
        A.~D\"oring\footnote{email: a.doering@imperial.ac.uk}\\[10pt]

\begin{center}                      and
\end{center}

        C.J.~Isham\footnote{email: c.isham@imperial.ac.uk}\\[20pt]

        The Blackett Laboratory\\ Imperial College of Science,
        Technology \& Medicine\\ South Kensington\\ London SW7 2BZ\\United
        Kingdom
\end{center}

\begin{center}
6 March, 2007
\end{center}

\begin{abstract}

This paper is the fourth in a series whose goal is to develop a
fundamentally new way of building theories of physics. The
motivation comes from a desire to address certain deep issues that
arise in the quantum theory of gravity. Our basic contention is
that constructing a theory of physics is equivalent to finding a
representation in a topos of a certain formal language that is
attached to the system. Classical physics arises when the topos is 
the category of sets. Other types of theory employ a
different topos.

The previous papers in this  series are concerned with
implementing this programme for a single system. In the present
paper, we turn to considering a \emph{collection} of systems: in
particular, we are interested in the relation between the topos
representation for a composite system, and the representations for
its constituents. We also study this problem for the disjoint sum
of two systems. Our approach to these matters is to construct a
\emph{category} of systems and to find a topos representation of
the entire category.

\end{abstract}
\end{titlepage}

\section{Introduction}
This is the fourth in a series of papers whose aim is to construct
a general framework within which theories of physics can be
expressed in a topos other than that of sets. In the first paper,
\cite{DI(1)}, we developed the idea of a typed, higher-order
(local) language, ${\cal L}_S$, for each physical system $S$, with
the goal of finding representations of this language in various
topoi. Then, in the second and third papers, \cite{DI(2), DI(3)},
we showed in detail how the `daseinisation' operation in quantum
theory enables us to represent this language---and a simpler
propositional language---in a certain topos of presheaves.

In the present paper, we  return to the more general aspects of
our theory, and study its application  to a \emph{collection} of
systems, each one of which may be associated with a  different
topos. For example, if $S_1,S_2$ is a pair of systems, with
associated topoi $\tau(S_1)$ and $\tau(S_2)$, and if $S_1$ is a
sub-system of $S_2$, then we wish to consider how $\tau(S_1)$ is
related to $\tau(S_2)$. Similarly, if a composite system is formed
from a pair of systems $S_1,S_2$, what relations are there between
the topos of the composite system and the topoi of the constituent
parts?

We start in Section  \ref{Sec:CatSys}, by introducing the notion
of a `category of systems', $\Sys$, whose objects are the physical
systems of interest, and whose arrows represent situations in
which one system is a `sub-system', or a `constituent' of another,
or combinations of such situations. A particular example of a
sub-system arises in the `disjoint sum' of two systems. We argue
on physical grounds that $\Sys$ can be regarded as a symmetric
monoidal category in two ways: one in which the monoidal product
represents forming a composite system, and one in which the
monoidal product represents the disjoint sum.  We show how such
arrows correspond to `translations' of the local languages
associated with the component systems. We also give a preliminary
definition of  the representation of the category $\Sys$ in a
category of topoi, ${\cal M}(\Sys)$. We then show that the scheme
works consistently in classical physics.

The idea of representing $\Sys$ is developed at length in Section
\ref{Sec:ToposAxioms}. An important ingredient is the `pull-back'
operation that arises when representing the arrows of $\Sys$ in
the category of topoi, ${\cal M}(\Sys)$.  Then, in Section
\ref{SubSec:GeneralToposAxioms} we bring together all these ideas
in the form of a set of rules for constructing a topos
representation of the objects and arrows in the Category $\Sys$.

In Section \ref{Sec:ReviewQT} we show how our earlier work (in
papers II and III) on toposifying quantum theory can be extended
to give a topos representation of $\Sys$. The disjoint sum of
systems behaves well under the pull-back operation but the
situation for the composition of systems is different: something,
we think, that reflects the existence of entanglement in quantum
theory. Finally, in Section \ref{Sec:conclusions} we speculate a
little on how the general scheme might be developed.

\section{The Category of Systems}
\label{Sec:CatSys}
\subsection{Background Remarks}
In one sense, there is  only one true `system', and that is the
universe as a whole. Concomitantly,  there  is just one local
language, and one topos. However, in practice,  the universe is
divided conceptually into portions that are sufficiently simple to
be amenable to theoretical discussion. Of course, this division is
not unique,  but it must be such that the coupling between
portions is weak enough that,  to a good approximation, their
theoretical models can be studied in isolation from each other.
Such an essentially isolated\footnote{The ideal monad has no
windows.} portion of the universe is called a `sub-system'. By an
abuse of language, sub-systems of the universe are usually called
`systems'  (so that the universe as a whole is one super-system),
and then we can  talk about `sub-systems' of these systems; or
`composites' of them; or sub-systems of the composite systems, and
so on.

Of course, in practice,  references by physicists to systems and
sub-systems\footnote{The word `sub-system' does not only mean a
collection of objects that is spatially localised. One could also
consider sub-systems of field systems by focussing on a just a few
modes of the fields as is done, for example, in the
Robertson-Walker model for cosmology. Another possibility would be
to use fields localised in some fixed space, or space-time region
provided that this is consistent with the dynamics.}   do not
generally signify \emph{actual} sub-systems of the real universe
but rather idealisations of possible systems. This is what a
physics lecturer  means when he or she starts a lecture by saying
``Consider a point particle moving in three dimensions.....''.

To develop these ideas further we need   mathematical control over
the  systems of interest, and their interrelations. To this end,
we start by focussing on some collection, $\Sys$, of physical
systems to which a particular theory-type is deemed to be
applicable. For example, we could consider   a collection of
systems that are to be discussed using the methodology of
classical physics; or systems to be discussed using standard
quantum theory; or whatever. For completeness, we  require that
every sub-system of a system in $\Sys$ is itself a member of
$\Sys$, as is every composite of members of $\Sys$.

We shall assume that  the systems in $\Sys$ are all associated
with local languages of the type discussed in paper I, and that
they all have the \emph{same} set of ground  symbols which, for
the purposes of the present discussion, we take to be just $\Si$
and $\R$. It follows that the languages $\L{S}$, $S\in\Sys$,
differ from each other only in the set of function symbols
$\F{S}$; \ie\  the set of \emph{physical quantities}.

As a simple example of the system-dependence of the set of
function symbols  let system $S_1$ be a point particle moving in
one dimension, and let the set of physical quantities be
$\F{S_1}=\{x,p,H\}$. In the language $\L{S_1}$, these
function-symbols represent the position, momentum, and energy of
the system respectively. On the other hand, if $S_2$ is a particle
moving in three dimensions, then in the language $\L{S_2}$ we
could have $\F{S_2}= \{x,y,z,p_x,p_y,p_z,H\}$ to allow for
three-dimensional position and momentum. Or, we could decide to
add angular momentum as well, to give the set $\F{S_2}=
\{x,y,z,p_x,p_y,p_z,J_x,J_y,J_z,H\}$.

\subsection{The Category $\Sys$}
\subsubsection{The Arrows and Translations for the Disjoint Sum
$S_1\sqcup S_2$.}\label{SubSubSec:ATDS}
 The use of local languages is central to our
overall topos scheme, and therefore we need to understand, in
particular, (i) the relation between the languages $\L{S_1}$ and
$\L{S_2}$ if $S_1$ is a sub-system of $S_2$; and (ii) the relation
between $\L{S_1}$, $\L{S_2}$ and $\L{S_1\di S_2}$, where $S_1\di
S_2$ denotes the composite of systems $S_1$ and $S_2$.

These discussions can be made more precise by regarding $\Sys$ as
a category whose objects are the systems.\footnote{To control the
size of  $\Sys$  we  assume that the collection of objects/systems
is  a \emph{set} rather than a more general class.} The arrows in
$\Sys$ need  to cover two basic types of relation: (i) that
between $S_1$ and $S_2$ if $S_1$ is a `sub-system' of $S_2$; and
(ii) that between a composite system, $S_1\di S_2$, and its
constituent systems, $S_1$ and $S_2$.

This may seem straightforward but, in fact, care is needed since
although the idea  of a `sub-system'  seems intuitively clear, it
is hard to give a physically acceptable definition that is
universal. However,  some insight into this idea can be gained by
considering its meaning in classical physics. This is very
relevant for the general scheme since one of our main goals is to
make all theories `look' like classical physics in the appropriate
topos.

To this end, let $S_1$ and $S_2$ be classical systems whose state
spaces are the symplectic manifolds $\S_1$ and $\S_2$
respectively. If $S_1$ is deemed to be a sub-system of $S_2$, it
is natural to require that $\S_1$ is a \emph{sub-manifold} of
$\S_2$, \ie\ $\S_1\subseteq\S_2$. However, this condition cannot
be used as a \emph{definition} of a `sub-system' since the
converse may not be true: \ie\ if $\S_1\subseteq\S_2$, this does
not necessarily mean that, from a physical perspective, $S_1$
could, or would, be said to be a sub-system of $S_2$.\footnote{
For example, consider the diagonal sub-manifold
$\De(\S)\subset\S\times\S$ of the symplectic manifold $\S\times\S$
that represents the composite $S\di S$ of two copies of a system
$S$. Evidently, the states in $\De(\S)$ correspond to the
situation in which both copies of $S$\ `march together'. It is
doubtful if this would be recognised physically as a sub-system.}

On the other hand, there are situations where being a sub-manifold
clearly \emph{does} imply being a physical sub-system. For
example, suppose the state space $\S$ of a system $S$ is a
disconnected manifold with two components $\S_1$ and $\S_2$, so
that $\S$ is the disjoint union, $\S_1\coprod\S_2$, of the
sub-manifolds $\S_1$ and $\S_2$.  Then it  seems physically
appropriate to say that the system $S$ itself is disconnected, and
to write $S=S_1\sqcup S_2$ where the symplectic manifolds that
represent the sub-systems $S_1$ and $S_2$ are $\S_1$ and $\S_2$
respectively.

One reason why it is reasonable to call $S_1$ and $S_2$
`sub-systems' in this particular situation is that any continuous
dynamical evolution of a state point in $\S\simeq\S_1\sqcup \S_2 $
will always lie in either one component or the other. This
suggests that perhaps, in general, a necessary condition for a
sub-manifold $\S_1\subseteq\S_2$ to represent a physical
sub-system is that the dynamics of the system $S_2$ must be such
that $\S_1$ is  mapped into itself  under the dynamical evolution
on $\S_2$; in other words, $\S_1$ is a
\emph{dynamically-invariant} sub-manifold of $\S_2$. This
correlates with the idea mentioned earlier that sub-systems are
weakly-coupled with each other.

However, such a dynamical restriction is not something that should
be coded into the languages, $\L{S_1}$ and $\L{S_2}$: rather, the
dynamics is to be  associated with the \emph{representation} of
these languages in the appropriate topoi.

Still, this caveat does not apply to the disjoint sum $S_1\sqcup
S_2$ of two systems $S_1,S_2$, and we will assume that, in
general, (\ie\ not just in classical physics) it is legitimate to
think of $S_1$ and $S_2$ as being sub-systems of $S_1\sqcup S_2$;
something that we indicate by defining  arrows $i_1:S_1\map
S_1\sqcup S_2$, and $i_2:S_2\map S_1\sqcup S_2$ in $\Sys$.

To proceed further it is important to understand the connection
between the putative arrows in the category $\Sys$, and the
`translations' of the associated languages. The first step is to
consider what can be said about the relation between $\L{S_1\sqcup
S_2}$, and $\L{S_1}$ and $\L{S_2}$. All three languages share the
same ground-type symbols, and so what we are concerned with is the
relation between the function symbols of signature $\Si\map\R$ in
these languages.

By considering what is meant intuitively by the disjoint sum, it
seems plausible that each physical quantity for the system
$S_1\sqcup S_2$ produces a physical quantity for $S_1$, and
another one for $S_2$. Conversely, specifying a pair of physical
quantities---one for $S_1$ and one for $S_2$---gives a physical
quantity for $S_1\sqcup S_2$. In other words,
\begin{equation}
        \F{S_1\sqcup S_2}\simeq \F{S_1}\times\F{S_2}\label{FS1sumS2}
\end{equation}
However, it is important not to be too dogmatic about statements
of this type since in non-classical theories  new possibilities
can arise that are counter to intuition.

Associated with \eq{FS1sumS2} are the maps $\L{i_1}:\F{S_1\sqcup
S_2}\map\F{S_1}$ and $\L{i_2}:\F{S_1\sqcup S_2}\map\F{S_2}$,
defined as  the projection maps of the product. In the theory of
local languages, these transformations are essentially
\emph{translations} \cite{Bell88} of $\L{S_1\sqcup S_2}$ in
$\L{S_1}$ and $\L{S_2}$ respectively; a situation that we denote
$\L{i_1}:\L{S_1\sqcup S_2}\map\L{S_1}$, and $\L{i_2}:\L{S_1\sqcup
S_2}\map\L{S_2}$.

To be more precise, these operations are translations if, taking
$\L{i_1}$ as the explanatory example, the map
$\L{i_1}:\F{S_1\sqcup S_2}\map \F{S_1}$ is supplemented with the
following map from the ground symbols of $\L{S_1\sqcup S_2}$ to
those of $\L{S_1}$:
\begin{eqnarray}
        \L{i_1}(\Si)&:=&\Si,            \label{tSigma}\\
        \L{i_1}(\R)&:=&\R,              \label{tR}    \\
        \L{i_1}(1)&:=&1,                \label{t1}    \\
        \L{i_1}(\Omega)&:=&\Omega.       \label{tOmega}
\end{eqnarray}
Such a translation map is then extended  to all type symbols using
the definitions
\begin{eqnarray}
        \L{i_1}(T_1\times T_2\times\cdots\times T_n)&=&
        \L{i_1}(T_1)\times \L{i_1}(T_2)  \times\cdots\times
        \L{i_1}(T_n), \\[2pt]
        \L{i_1}(PT)&=&P[\L{i_1}(T)]
\end{eqnarray}
for all finite $n$ and all type symbols $T,T_1,T_2,\ldots,T_n$.
This, in turn, can be extended inductively to all terms in the
language. Thus, in our case, the translations act trivially on all
the type symbols.

\paragraph{Arrows in $\Sys$ \emph{are} translations.}
Motivated by this argument  we now turn everything around and, in
general, \emph{define} an arrow $j:S_1\map S$ in the category
$\Sys$ to mean that there is some \emph{physically meaningful} way
of transforming the physical quantities in $S$ to physical
quantities in $S_1$. If, for any pair of systems $S_1,S$ there is
more than one such transformation, then there will be more than
one arrow from $S_1$ to $S$.

To make this more precise, let $\Loc$ denote the collection of all
(small\footnote{This means that the collection of symbols is a
set, not a more general class.}) local languages. This is a
category whose objects are the local languages, and whose arrows
are translations between languages.  Then our basic assumption is
that the association $S\mapsto\L{S}$ is a covariant functor from
$\Sys$ to $\Loc^{\rm op}$, which we denote as ${\cal
L}:\Sys\map\Loc^{\rm op}$.

Note that the  combination of a pair of arrows in $\Sys$  exists
in so far as the associated translations can be combined.

\subsubsection{The Arrows and Translations for the
Composite System $S_1\di S_2$.}\label{SubSubSec:ArrTranComp} Let
us now consider the composition $S_1\di S_2$ of a pair of systems.
In the case of classical physics,  if $\S_1$ and $\S_2$ are the
symplectic manifolds that represent the systems $S_1$ and $S_2$
respectively, then the manifold that represents the composite
system is the cartesian product $\S_1\times\S_2$. This is
distinguished by the existence of the two projection functions
$\pr_1:\S_1\times \S_2\map \S_1$ and $\pr_2:\S_1\times
\S_2\map\S_2$.

It seems reasonable to impose the same type of structure on
$\Sys$: \ie\ to require there to be arrows $p_1:S_1\di S_2\map
S_1$ and $p_2:S_1\di S_2\map S_2$ in $\Sys$. However, bearing in
mind the definition above, these arrows $p_1,p_2$ exist if, and
only if, there are corresponding translations
$\L{p_1}:\L{S_1}\map\L{S_1\di S_2}$, and
$\L{p_2}:\L{S_2}\map\L{S_1\di S_2}$. But there \emph{are} such
translations: for if $A_1$ is a physical quantity for system
$S_1$, then $\L{p_1}(A_1)$ can be defined as that same physical
quantity, but now regarded as pertaining to the combined system
$S_1\di S_2$; and analogously for system $S_2$.\footnote{For
example, if $A$ is the energy of particle $1$, then we can talk
about this energy in the combination of a pair of particles. Of
course, in---for example---classical physics there is no reason
why the energy of particle $1$ should be \emph{conserved} in the
composite system, but that, dynamical, question is a different
matter.}  We shall denote this translated quantity,
$\L{p_1}(A_1)$, by $A_1\di 1$.

Note that we do \emph{not} postulate any simple relation between
$\F{S_1\di S_2}$ and $\F{S_1}$ and $\F{S_2}$; \ie\ there is no
analogue of \eq{FS1sumS2} for combinations of systems.

The definitions above of the basic arrows suggest that we might
also want to impose the following conditions:
\begin{enumerate}
\item The arrows $i_1:S_1\map S_1\sqcup S_2$, and
$i_2:S_2\map S_1\sqcup S_2$ are \emph{monic} in $\Sys$.

\item The arrows $p_1:S_1\di S_2\map S_1$ and
$p_2:S_1\di S_2\map S_2$ are \emph{epic} arrows in $\Sys$.
\end{enumerate}
However, we do \emph{not} require that $S_1\cup S_2$ and $S_1\di
S_2$ are the co-product and product, respectively, of $S_1$ and
$S_2$ in the category $\Sys$.

\subsubsection{The Concept of `Isomorphic' Systems.}
We also need to decide  what it means to say that two systems
$S_1$ and $S_2$ are \emph{isomorphic}, to be denoted $S_1\simeq
S_2$. As with the concept of sub-system, the notion of isomorphism
is to some extent a matter of definition rather than obvious
physical structure, albeit with the expectation that isomorphic
systems in $\Sys$ will correspond to isomorphic local languages,
and be represented by isomorphic mathematical objects in any
concrete realisation of the axioms: for example, by isomorphic
symplectic manifolds in classical physics.

To a considerable extent, the physical meaning of `isomorphism'
depends on whether one is dealing with actual physical systems, or
idealisations of them. For example,  an electron confined in a box
in Cambridge is presumably isomorphic to one  confined in the same
type of box in London, although they are not the same physical
system. On the other hand, when a lecturer says ``Consider an
electron trapped in a box....'', he/she is referring to an
idealised system.

One could, perhaps, say that an idealised system is an
\emph{equivalence class} (under isomorphisms)\ of real systems,
but even working only with  idealisations does not entirely remove
the need for the concept of isomorphism.

For example, in classical mechanics, consider the (idealised)\
system $S$ of a point particle moving in a box, and let $1$ denote
the `trivial system' that consists of just a single  point with no
internal or external degrees of freedom. Now consider the  system
$S\di 1$. In classical mechanics this is represented by the
symplectic manifold $\S\times\{*\}$, where $\{*\}$\ is a single
point, regarded as a zero-dimensional manifold. However,
$\S\times\{*\}$ is isomorphic to the manifold $\S,$ and it is
clear physically that the system $S\di 1$ is isomorphic to the
system $S$.  On the other hand, one cannot say that $S\di 1$ is
literally \emph{equal} to $S$, so the concept of `isomorphism'
needs to be maintained.

One thing that \emph{is} clear is that if $S_1\simeq S_2$ then
$\F{S_1}\simeq \F{S_2}$, and if any other non-empty sets of
function symbols are present, then they too must be isomorphic.

Note that when introducing a trivial system, $1$, it necessary to
specify its local language, $\L{1}$. The set of function symbols
$\F{1}$ is not completely empty since, in classical physics, one
does have a preferred physical quantity, which is just the number
$1$. If one asks what is meant in general by the `number $1$' the
answer is not trivial since, in the reals $\mathR$, the number $1$
is the multiplicative identity. It would be possible to add the
existence of such a unit to the axioms for $\R$ but this would
involve introducing a multiplicative structure and we do not know
if there might be physically interesting topos representations
that do not have this feature.

For the moment then, we will  say that the trivial system has just
a single physical quantity, which in classical physics translates
to the number $1$.  More generally, for the language $\L{1}$ we
specify that $\F{1}:=\{I\}$, \ie\ $\F{1}$ has just a single
element, $I$, say. Furthermore, we add the axiom
\begin{equation}
 :\forall \va{s}_1\forall \va{s}_2,I(\va{s}_1)=I(\va{s}_2),
\end{equation}
where $\va{s}_1$ and $\va{s}_2$ are variables of type $\Si$. In
fact, it seems natural to add such a trivial quantity to the
language $\L{S}$ for \emph{any} system $S$, and from now on we
will assume that this has been done.

A related issue is that, in classical physics, if $A$ is a
physical quantity, then so is $rA$ for any $r\in\mathR$. This is
because the set of classical quantities
$A_\s:\Si_\s\map\R_\s\simeq\mathR$ forms a \emph{ring} whose
structure derives from the ring structure of $\mathR$. It would be
possible to add ring axioms for $\R$ to the language $\L{S}$, but
we think this is too strong, not least because, as shown in paper
III, it fails in quantum theory \cite{DI(3)}. Clearly, the general
question of axioms for $\R$ needs more thought: a task for later
work.

If desired, an `\emph{empty}' system, $0$, can be added too, with
$\F{0}:=\emptyset$. This, so called, `pure language', $\L{0}$, is
an initial object in the category $\Loc$.

\subsubsection{An Axiomatic Formulation of $\Sys$}
Let us now  summarise, and clarify, our list of axioms for a
category $\Sys$:
\begin{enumerate}
        \item  The collection $\Sys$ is a small category where
    (i) the objects are the systems of interest (or, if desired,
    isomorphism classes of such systems); and (ii)
    the arrows are defined as above.

        Thus the fundamental property of an arrow $j:S_1\map S$ in $\Sys$
        is that it induces, and is essentially \emph{defined by}, a
        translation $\L{j}:\L{S}\map \L{S_1}$. Physically, it corresponds
        to  the physical quantities for system $S$ being `pulled-back' to
        give physical quantities for system $S_1$.

        Arrows of particular interest are those associated with
        `sub-systems' and `composite systems', as discussed above.

        \item The axioms for a category are satisfied because:
        \begin{enumerate}
            \item Physically, the ability to form composites of
            arrows follows from the concept of `pulling-back' physical
            quantities. From a mathematical perspective,
            if $j:S_1\map S_2$ and $k:S_2\map S_3$,
            then the translations give functions $\L{j}:\F{S_2}
                \map \F{S_1}$ and $\L{k}:\F{S_3} \map \F{S_2}$. Then clearly
                $\L{j}\circ \L{k}:\F{S_3}\map \F{S_1}$, and this can thought of as
                the translation corresponding to the arrow $k\circ j:S_1\map S_3$.

            The associativity of the law of arrow combination can be
            proved in a similar way.

           \item We add by hand a special arrow ${\rm id}_S : S \map S$
           which is defined to correspond to the translation $\L{\id_S}$ that
          is given by the identity map on $\F{S}$. Clearly, ${\rm id}_S :
         S \map S$ acts an an identity morphism should.
        \end{enumerate}

        \item For any pair of systems $S_1,S_2$, there is a
        \emph{disjoint sum}, denoted $S_1\sqcup S_2$. The disjoint sum has
        the following properties:
        \begin{enumerate}
            \item For all systems $S_1,S_2,S_3$ in $\Sys$:
            \begin{equation}
                        (S_1\sqcup S_2)\sqcup S_3\simeq S_1\sqcup (S_2\sqcup S_3).
            \end{equation}

            \item For all systems $S_1, S_2$ in $\Sys$:
        \begin{equation}
                    S_1\sqcup S_2 \simeq S_2\sqcup S_1.
            \end{equation}

        \item There are  arrows in $\Sys$:
        \begin{equation}
                        i_1:S_1\map S_1\sqcup S_2 \mbox{\ \ and }
                        i_2:S_2\map S_1\sqcup S_2
        \end{equation}
          that are associated with translations in the sense discussed in
          Section \ref{SubSubSec:ATDS}. These are associated with the
          decomposition
                \begin{equation}
                \F{S_1\sqcup S_2}\simeq\F{S_1}\times\F{S_2}.
                                \label{F(S1+S2)=FS1xFS2}
                \end{equation}
        \end{enumerate}

        We assume that if $S_1,S_2$ belong to $\Sys$, then $\Sys$ also
        contains $S_1\sqcup S_2$.

        \item For any given pair of systems $S_1,S_2$, there is a
        \emph{composite} system in $\Sys$, denoted\footnote{The product
        operation in a monoidal category is often written `$\otimes$'.
        However,  a different symbol  has been used here to avoid
        confusion with existing usages in physics of the tensor product
        sign `$\otimes$'.} $S_1\di S_2$, with the following properties:
        \begin{enumerate}
            \item For all systems $S_1,S_2,S_3$ in $\Sys$:
            \begin{equation}
                        (S_1\di S_2)\di S_3 \simeq
                 S_1\di (S_2\di S_3).\label{(Sys: Assoc)}
            \end{equation}

            \item For all systems $S_1, S_2$ in $\Sys$:
        \begin{equation}
                    S_1\di S_2 \simeq S_2 \di S_1.            \label{(SysSym)}
            \end{equation}

            \item There are  arrows in $\Sys$:
        \begin{equation}
                        p_1:S_1\di S_2 \map S_1\mbox{ and }
                        p_2:S_1\di S_2 \map S_2
        \end{equation}
          that are associated with translations in the sense discussed
           in Section \ref{SubSubSec:ArrTranComp}.
        \end{enumerate}
        We assume that if $S_1,S_2$ belong to $\Sys$, then $\Sys$ also
        contains the composite system $S_1\di S_2$.

        \item It seems physically reasonable to add the axiom
        \begin{equation}
            (S_1\sqcup S_2)\di S\simeq (S_1\di S)\sqcup (S_2\di S)
                        \label{S1cupS2diS}
        \end{equation}
        for all systems $S_1,S_2,S$. However, physical intuition can
        be a dangerous thing, and so, as with most of these axioms,
        we are not dogmatic, and feel free to change them as
        new insights emerge.

        \item There is a trivial system, $1$,  such that for all systems
        $S$, we have
        \begin{equation}
        S \di 1\simeq  S \simeq1 \di S
                                \label{(Sys: Unit)}
        \end{equation}

        \item It may be convenient to postulate an `empty system', $0$,
        with the properties
        \begin{eqnarray}
        S\di 0&\simeq& 0\di S\simeq 0 \\
        S\sqcup 0&\simeq& 0\sqcup S\simeq S
        \end{eqnarray}
        for all systems $S$.

        Within the meaning given to arrows in $\Sys$, $0$ is a
        \emph{terminal object} in $\Sys$. This is because the empty set of
        function symbols of signature $\Si\map\R$ is a subset of any other
        set of function symbols of this signature.
\end{enumerate}

It might seem tempting to postulate that composition laws are
well-behaved with respect to arrows. Namely, if $j:S_1\map S_2$,
then, for any $S$, there is an arrow $S_1\di S\map S_2\di S$ and
an arrow $S_1\sqcup S\map S_2\sqcup S$.\footnote{A more accurate
way of capturing this idea is to say that the operation
$\Sys\times\Sys\map\Sys$ in which
\begin{equation}
\langle S_1,S_2\rangle\mapsto S_1\di S_2 \label{(Sys: BiFunc)}
\end{equation}
is a \emph{bi-functor} from $\Sys\times\Sys$ to $\Sys$. Ditto for
the operation in which $\la S_1,S_2\ra\mapsto S_1\sqcup S_2$.}

In the case of the disjoint sum, such an arrow can be easily
constructed using \eq{F(S1+S2)=FS1xFS2}. First split the function
symbols in $\F{S_1\sqcup S}$ into $\F{S_1}\times\F{S}$ and the
function symbols in $\F{S_2\sqcup S}$ into $\F{S_2}\times\F{S}$.
Since there is an arrow $j:S_1\map S_2$, there is a translation
$\L{j}:\L{S_2}\map \L{S_1}$, given by a mapping
$\L{j}:\F{S_2}\map\F{S_1}$. Of course, then there is also a
mapping $\L{j}\times
\L{\id_S}:\F{S_2}\times\F{S}\map\F{S_1}\times\F{S}$, \ie\ a
translation between $\L{S_2\sqcup S}$ and $\L{S_1\sqcup S}$. Since
we assume that there is an arrow in $\Sys$ whenever there is a
translation (in the opposite direction), there is indeed  an arrow
$S_1\sqcup S\map S_2\sqcup S$.

In the case of the composition, however, this would require a
translation $\L{S_2\di S}\map \L{S_1\di S}$, and this cannot be
done in general since we have no \emph{prima facie} information
about the  set of function symbols $\F{S_2\di S}$. However, if we
restrict the arrows in $\Sys$ to be those associated with
sub-systems, combination of systems, and compositions of such
arrows, then it is easy to see that the required translations
exist (the proof of this makes essential use of \eq{S1cupS2diS}).

If we make this restriction of arrows, then the axioms
\eq{(SysSym)}, \eqs{(Sys: Unit)}{(Sys: BiFunc)}, mean that,
essentially, $\Sys$ has the structure of a \emph{symmetric
monoidal}\footnote{In the actual definition of a monoidal category
the two isomorphisms in \eq{(Sys: Unit)} are separated from each
other, whereas  we have identified them. Further more, these
isomorphism are required to be natural. This seems a correct thing
to require in our case, too.} category in which the monoidal
product operation is `$\di$', and the left and right unit object
is $1$. There is also a monoidal structure associated with the
disjoint sum `$\sqcup$', with $0$ as the unit object.

We say `essentially' because in order to comply with all the
axioms of a monoidal category, $\Sys$ must satisfy certain
additional, so-called, `coherence' axioms. However, from a
physical perspective these are very plausible statements about (i)
how the unit object $1$ intertwines with the $\di$-operation; how
the null object intertwines with the $\sqcup$-operation; and (iii)
certain properties of quadruple products (and disjoint sums) of
systems.

\paragraph{A simple example of a category $\Sys$.}
It might be helpful at this point to give a simple example of a
category $\Sys$. To that end, let $S$ denote a point particle that
moves in three dimensions, and let us suppose that $S$ has no
sub-systems other than the trivial system $1$. Then  $S\di S$ is
defined to be a pair of particles moving in three dimensions, and
so on. Thus the objects in our category are $1$, $S$, $S\di S$,
$\ldots$,  $S\di S\di\cdots S$ $\ldots$ where the `$\di$'
operation is formed any finite number of times.

At this stage, the only arrows are those that are associated with
the constituents of a composite system. However, we could
contemplate adding to the systems the disjoint sum $S\sqcup (S\di
S)$ which is a system that is either one particle or two particles
(but, of course, not both at the same time). And, clearly, we
could extend this to $S\sqcup (S\di S)\sqcup (S\di S\di S)$, and
so on. Each of these disjoint sums comes with its own arrows, as
explained above.

Note that this particular category of systems has the property
that it can be treated using either classical physics or quantum
theory.

\subsection{Representations of $\Sys$ in Topoi}
We assume that all the systems in $\Sys$ are to be treated with
the same theory type. We also assume that systems in $\Sys$ with
the \emph{same} language are to be represented in the same topos.
Then we define:\footnote{As emphasised already, the association
$S\mapsto \L{S}$ is generally not one-to-one: \ie\ many systems
may share the same language. Thus, when we come discuss the
representation of the language $\L{S}$ in a topos, the extra
information about the system $S$  is used in fixing the
representation.} {\definition\label{Defn:TopReal} A \emph{topos
realisation} of $\Sys$ is an association, $\phi$, to each system
$S$ in $\Sys$, of a triple $\phi(S)=\la\rho_{\phi,
S},\L{S},\tau_\phi(S)\ra$ where:

\begin{enumerate}
        \item[(i)] $\tau_\phi(S)$ is the topos in which the
        theory-type applied to system $S$ is to be realised.

       \item[(ii)] $\L{S}$ is the local language in $\Loc$
       that is associated with $S$.
        This is not dependent on the realisation $\phi$.

\item[(iii)]$\rho_{\phi, S}$ is
       a representation of the local language $\L{S}$ in the
       topos $\tau_\phi(S)$. As a more descriptive piece of notation
        we write $\rho_{\phi,S}:\L{S}\sq\tau_\phi(S)$.
        The key part of this representation is the map
\begin{equation}
        \rho_{\phi,S}:\F{S}\map\Hom{\tau_\phi(S)}{\Si_{\phi,S}}
        {\R_{\phi,S}}
\end{equation}
where $\Si_{\phi,S}$ and $\R_{\phi,S}$ are the state object and
quantity-value object, respectively, of the representation $\phi$
in the topos $\tau_\phi(S)$. As a convenient piece of notation we
write $A_{\phi,S}:=\rho_{\phi,S}(A)$ for all $A\in\F{S}$.
\end{enumerate}
} \noindent This definition is only partial; the possibility of
extending it will be discussed shortly.

Now, if $j:S_1\map S$ is an arrow in $\Sys$, then there is a
translation arrow ${\cal L}{(j)}:\L{S}\map\L{S_1}$. Thus we have
the beginnings of a commutative diagram
\begin{equation}
\setsqparms[1`1`-1`1;1000`700]\label{ComDiag}
\square[S_1`\la\rho_{\phi,S_1},\L{S_1},\tau_\phi(S_1)
\ra`S`\la\rho_{\phi,S},\L{S},\tau_\phi(S)\ra; \phi`j`?\times{\cal
L}(j)\times?`\phi]
\end{equation}
However, to be useful, the arrow on the right hand side of this
diagram should refer to some relation between (i) the topoi
$\tau_\phi(S_1)$ and $\tau_\phi(S)$; and (ii) the realisations
$\rho_{\phi,S_1}:\L{S_1}\sq\tau_\phi(S_1)$ and
$\rho_{\phi,S}:\L{S}\sq\tau_\phi(S)$: this is the significance of
the two `?' symbols in the arrow written `$?\times\L{j}\times?$'.

Indeed, as things stand, Definition \ref{Defn:TopReal} says
nothing  about relations between the topoi representations of
different systems in $\Sys$. We are particularly interested in the
situation where there are two different systems $S_1$ and $S$ with
an arrow $j:S_1\map S$ in $\Sys$.

We know that the arrow $j$ is associated with a translation
$\L{j}:\L{S}\map\L{S_1}$, and an attractive possibility,
therefore, would be to seek, or postulate, a `covering' map
$\phi(\L{j}): \Hom{\tau_\phi(S)}{\Si_{\phi,S}}{\R_{\phi,S}} \map
\Hom{\tau_\phi(S_1)}{\Si_{\phi,S_1}}{\R_{\phi,S_1}}$ to be
construed as a topos representation of the translation
$\L{j}:\L{S}\map \L{S_1}$, and hence of the  arrow $j:S_1\map S$
in $\Sys$.

This raises the questions of what properties these `translation
representations' should possess in order to justify saying that
they `cover' the translations. A minimal requirement is that if
$k:S_2\map S_1$ and $j:S_1\map S$, then the map $\phi(\L{j\circ
k}): \Hom{\tau_\phi(S)}{\Si_{\phi,S}}{\R_{\phi,S}} \map
\Hom{\tau_\phi(S_2)}{\Si_{\phi,S_2}}{\R_{\phi,S_2}}$ factorises as
\begin{equation}
        \phi(\L{j\circ k})=\phi(\L{k})\circ\phi(\L{j}).
                                                \label{phiL(jcirck)}
\end{equation}
We also require that
\begin{equation}
\phi(\L{\id_S})=\id: \Hom{\tau_\phi(S)}{\Si_{\phi,S}}{\R_{\phi,S}}
\map \Hom{\tau_\phi(S)}{\Si_{\phi,S}}{\R_{\phi,S}}
\label{phiL(idS)}
\end{equation}
for all systems $S$.

The conditions \eq{phiL(jcirck)} and \eq{phiL(idS)} seem eminently
plausible, and they are not particularly strong. A far more
restrictive axiom would be to require the following diagram to
commute:
\begin{equation}\label{ComDLphi0}
        \setsqparms[1`1`1`1;1200`700]
        \square[\F{S}`\Hom{\tau_\phi(S)}{\Si_{\phi,S}}
        {\R_{\phi,S}}`\F{S_1}`\Hom{\tau_\phi(S_1)}
        {\Si_{\phi,S_1}}{\R_{\phi,S_1}};
        \rho_{\phi,S}`\L{j}`\phi(\L{j})`\rho_{\phi,S_1}]
\end{equation}
At first sight, this requirement seems very appealing. However,
caution is needed when postulating `axioms' for a theoretical
structure in physics. It is easy to get captivated by the
underlying mathematics and to assume, erroneously, that  what is
mathematically elegant is necessarily true in the physical theory.

The translation $\phi(\L{j})$ maps an arrow from $\Si_{\phi,S}$ to
$\R_{\phi,S}$ to an arrow from $\Si_{\phi,S_1}$ to
$\R_{\phi,S_1}$. Intuitively, if $\Si_{\phi,S_1}$ is a `much
larger' object than $\Si_{\phi,S}$ (they lie in different topoi,
so no direct comparison is available), the translation can only be
`faithful' on some part of $\Si_{\phi,S_1}$ that can be identified
with (the `image' of) $\Si_{\phi,S}$. A concrete example of this
will show up in the treatment of composite quantum systems, see
Subsection \ref{SubSec:TranslCompQT}. As one might expect, a form
of entanglement plays a role here.

\subsection{Classical Physics in This  Form}
\subsubsection{The Rules so Far.}
Constructing maps $\phi(\L{j}):
\Hom{\tau_\phi(S)}{\Si_{\phi,S}}{\R_{\phi,S}} \map
\Hom{\tau_\phi(S_1)}{\Si_{\phi,S_1}}{\R_{\phi,S_1}}$ is likely to
be complicated when $\tau_\phi(S)$ and $\tau_\phi(S_1)$ are
different topoi, and so we begin  with the example of classical
physics, where the topos is always $\Set$.

In general, we are interested in the relation(s) between the
representations $\rho_{\phi,S_1}: \L{S_1}\sq\tau_\phi(S_1)$ and
$\rho_{\phi,S}:\L{S}\sq\tau_\phi(S)$ that is associated with an
arrow $j:S_1\map S$ in $\Sys$. In classical physics, we only have
to study the relation between the representations $\rho_{\s,S_1}:
\L{S_1}\sq\Set$ and $\rho_{\s,S}:\L{S}\sq\Set$.

Let us summarise what we have said so far (with $\s$ denoting the
$\Set$-realisation of classical physics):
\begin{enumerate}
    \item For any system $S$ in $\Sys$, a representation
        $\rho_{\s,S}:\L{S}\sq\Set$ consists of the following ingredients.
        \begin{enumerate}
                \item The ground symbol $\Si$  is
                represented by a symplectic  manifold,
                $\Si_{\s,S}:=\rho_{\s,S}(\Si)$, that serves as the
                classical state space.

                \item For all systems $S$, the ground symbol $\cal R$ is
                represented by the real numbers $\mathR$, \ie\
                $\R_{\s,S}=\mathR$, where $\R_{\s,S}:=\rho_{\s,S}(\R)$.

                \item Each function symbol $A:\Si\map\R$ in $\F{S}$ is
                represented by a function $A_{\s,S}=\rho_{\s,S}(A):
                \Si_{\s,S}\map\mathR$ in the set of functions\footnote{In
                practice, these functions are required to be measurable
                with respect to the Borel structures on the symplectic
                manifold $\Si_\s$ and $\mathR$. Many of the functions
                will also be smooth, but we will not go into such
                details here.} $C(\Si_{\s,S}; \mathR)$.
        \end{enumerate}

        \item The trivial system is mapped to a singleton set $\{*\}$
    (viewed as a zero-dimensional symplectic manifold):
    \begin{equation}
            \Si_{\s,1} := \{*\}.
    \end{equation}
        The empty system is represented by the empty set:
        \begin{equation}
        \Si_{\s,0}:=\emptyset.
        \end{equation}

        \item Propositions about the system $S$ are represented by (Borel)
        subsets of the state space $\Si_{\s,S}$.

        \item The composite system $S_1\di S_2$
        is represented by the
    Cartesian product $\Si_{\s,S_1}\times \Si_{\s,S_2}$; \ie\
    \begin{equation}
            \Si_{\s,\,S_1\di S_2}\simeq \Si_{\s,S_1}\times\Si_{\s,S_2}.
                                            \label{sigmaS1S2}
    \end{equation}

    The disjoint sum $S_1\sqcup S_2$ is represented by the disjoint
        union $\Si_{\s,S_1}\coprod \Si_{\s,S_2}$;\ie\
    \begin{equation}
        \Si_{\s,S_1\sqcup S_2}\simeq\Si_{\s,S_1}{\textstyle\coprod}\Si_{\s,S_2}.
     \end{equation}

    \item Let $j : S_1\map S$ be an arrow in $\Sys$. Then
        \begin{enumerate}
            \item There is a  translation map $\L{j}:\F{S}\map \F{S_1}$.

                \item  There is a symplectic function $\s(j):\Si_{\s,S_1}
        \map \Si_{\s,S}$ from the symplectic manifold $\Si_{\s,S_1}$
                to the symplectic manifold $\Si_{\s,S}$.
    \end{enumerate}
\end{enumerate}

The existence of this function $\s(j):\Si_{\s,S_1} \map
\Si_{\s,S}$ follows directly from the properties of sub-systems
and composite systems in classical physics. It is discussed in
detail below in Section \eq{SubSubSec:DetailsTrans}. As we shall
see, it underpins the classical realisation of our axioms.

These properties of the arrows stem from the fact that the
linguistic function symbols in $\F{S}$ are represented by
real-valued functions in $C(\Si_{\s,S},\mathR)$. Thus we can write
$\rho_{\s,S}:\F{S}\map C(\Si_{\s,S},\mathR)$, and similarly
$\rho_{\s,S_1}:\F{S_1}\map C(\Si_{\s,S_1},\mathR)$. The diagram in
\eq{ComDLphi0} now becomes
\begin{equation}\label{ComDLsClass}
        \setsqparms[1`1`1`1;1000`700]
        \square[\F{S}`C(\Si_{\s,S},\mathR)`\F{S_1}`C(\Si_{\s,S_1},\mathR);
        \rho_{\s,S}`\L{j}`\s(\L{j})`\rho_{\s,S_1}]
\end{equation}
and, therefore, the question of interest is if there is a
`translation representation' function
$\s(\L{j}):C(\Si_{\s,S},\mathR)\map C(\Si_{\s,S_1},\mathR)$ so
that this diagram commutes.

Now, as stated above, a physical quantity, $A$, for the system $S$
is represented in classical physics by a real-valued function
$A_{\s,S}=\rho_{\s,S}(A) :\Si_{\s, S}\map\mathR$. Similarly, the
representation of $\L{j}(A)$  for $S_1$ is given by a function
$A_{\s,S_1}:=\rho_{\s,S_1}(A):\Si_{\s,S_1}\map \mathR$. However,
in this classical case we also have the function
$\s(j):\Si_{\s,S_1}\map \Si_{\s,S}$, and it is clear that we can
use it to define
$[\rho_{\s,S_{1}}(\L{j}(A)](s):=\rho_{\s,S}(A)\big(\s(j)(s)\big)$
for all $s\in\Si_{\s,S_1}$. In other words
\begin{equation}
        \rho_{\s,S_1}\big(\L{j}(A)\big)=\rho_{\s,S}(A)\circ\s(j)
\end{equation}
or, in simpler notation
\begin{equation}
        \big((\L{j}(A)\big)_{\s,S_1}=A_{\s,S}\circ\s(j).
\end{equation}
But then it is clear that a translation-representation function
$\s(\L{j}):C(\Si_{\s,S},\mathR) \map C(\Si_{\s,S_1},\mathR)$ with
the desired property of making \eq{ComDLsClass} commute can be
defined by
\begin{equation}
\s(\L{j})(f):=f\circ\s(j)       \label{ClassPB}
\end{equation}
for all $f\in C(\Si_{\s,S},\mathR)$; \ie\ the function
$\s(\L{j})(f):\Si_{\s,S_1}\map\mathR$ is the usual pull-back of
the function $f:\Si_{\s,S}\map\mathR$ by the function
$\s(j):\Si_{\s,S_1}\map\Si_{\s,S}$. Thus, in the case of classical
physics, the commutative diagram in \eq{ComDiag} can be completed
to give
\begin{equation}                                \label{ComDiagCl}
        \setsqparms[1`1`-1`1;1000`700]
        \square[S_1`\la\rho_{\s,S_1},\L{S_1},
        \Set\ra`S`\la\rho_{\s,S},\L{S},\Set\ra;
        \s`j`\s(\L{j})\times\L{j}\times\id`\s]
\end{equation}

\subsubsection{Details of the Translation Representation.}
\label{SubSubSec:DetailsTrans}
\paragraph{The translation representation for a disjoint sum of classical
systems.} We first consider arrows of the form
\begin{equation}
S_{1}\overset{i_{1}}{\map}S_{1}\sqcup S_{2}\overset{i_{2}
}{\leftarrow}S_{2}
\end{equation}
from the components $S_{1}$, $S_{2}$ to the disjoint sum
$S_{1}\sqcup S_{2}$. The systems $S_{1}$, $S_{2}$ and $S_{1}\sqcup
S_{2}$ have symplectic manifolds $\Si_{\s,S_{1}}$,
$\Si_{\s,S_{2}}$ and $\Si_{\s,S_{1}\sqcup
S_{2}}=\Si_{\s,S_{1}}\coprod\Si_{\s,S_{2}}$. We write $i:=i_{1}$.

Let $S$ be a classical system. We assume that the function symbols
$A\in \F{S}$ in the language $\L{S}$ are in bijective
correspondence with an appropriate subset of the functions
$A_{\s,S}\in C(\Si_{\s,S},\mathR)$.\footnote{Depending on the
setting, one can assume that $\F{S}$ contains function symbols
corresponding bijectively to measurable, continuous or smooth
functions.}

There is an obvious translation representation. For if
$A\in\F{S_1\sqcup S_2}$, then since $\Si_{\s,S_1\sqcup
S_2}=\Si_{\s,S_1}\coprod\Si_{\s,S_1}$, the associated function
$A_{\s, S_1\sqcup S_2}:\Si _{\s,S_1\sqcup S_2} \map\mathR$ is
given by a pair of functions $A_{1}\in C(\Si_{\s,S_1},\mathR)$ and
$A_{2}\in C(\Si_{\s,S_2},\mathR)$; we write $A_{\s, S_1\sqcup
S_2}=\la A_1,A_2\ra$. It is natural to demand that the translation
representation $\s(\L{i})(A_{\s, S_1\sqcup S_2})$ is $A_1$.  Note
that what is essentially being discussed here is the
classical-physics representation of the relation \eq{FS1sumS2}.

The canonical choice for $\s(i)$ is
\begin{eqnarray}
  \s(i):\Si_{\s,S_1}  &  \map&\Si_{\s,S_1\sqcup S_2}=
                         \Si_{\s,S_1}{\textstyle\coprod}\Si_{\s,S_2}\\
                s_{1}  &  \mapsto& s_1.
\end{eqnarray}
Then the pull-back along $\s(i)$,
\begin{eqnarray}
\s(i)^*:C(\Si_{\s,S_1\sqcup S_2},\mathR)  & \map& C(\Si_{\s,S_1},\mathR)\\
A_{\s,S_1\sqcup S_2}  &  \mapsto& A_{\s,S_1\sqcup S_2} \circ
\s(i),
\end{eqnarray}
maps (or `translates') the topos representative $A_{\s,S_1\sqcup
S_2} =\la A_1, A_2\ra$ of the function symbol $A\in\F{S_1\sqcup
S_2}$ to a real-valued function $A_{\s,S_1\sqcup S_2} \circ \s(i)$
on $\Si_{\s,S_1}$. This function is clearly equal to $A_1$.

\paragraph{The translation in the case of a composite classical system.}
We now consider arrows in $\mathbf{Sys}$ of the form
\begin{equation}
S_{1}\overset{p_{1}}{\leftarrow}S_{1}\di S_{2}\overset{p_2
}{\map}S_2
\end{equation}
from the composite classical system $S_1\di S_2$ to the
constituent systems $S_1$ and $S_2$. Here, $p_1$ signals that
$S_{1}$ is a constituent of the composite system $S_1\di S_2$,
likewise $p_2$. The systems $S_{1}$, $S_{2}$ and $S_{1}\di S_{2}$
have symplectic manifolds $\Si_{\s,S_{1}}$, $\Si_{\s,S_{2}}$ and
$\Si_{\s,S_{1}\di S_{2} }=\Si_{\s,S_{1}}\times\Si_{\s,S_{2}}$,
respectively; \ie\  the state space of the composite system
$S_{1}\di S_{2}$ is the cartesian product of the state spaces of
the components. For typographical simplicity in what follows we
denote $p:=p_{1}$.

There is a canonical translation $\L{p}$ between the languages
$\L{S_1}$ and $\L{S_{1}\di S_{2}}$ whose representation is the
following. Namely, if $A$ is in $\F{S_1}$, then the corresponding
function $A_{\s,S_1}\in C(\Si_{\s,S_{1} },\mathR)$ is translated
to a  function $\s(\L{p}) (A_{\s,S_1})\in C(\Si_{\s,S_{1}\di
S_{2}},\mathR)$ such that
\begin{equation}
        \s(\L{p})(A_{\s,S_1})(s_1,s_2)=A_{\s,S_1}(s_1)
\end{equation}
for all $(s_1,s_2)\in\Si_{\s,S_1}\times\Si_{\s,S_2}$.

This natural translation representation is based on the fact that,
for the symplectic manifold $\Si_{\s,S_1\di
S_2}=\Si_{\s,S_1}\times\Si_{\s,S_2}$, each point
$s\in\Si_{\s,S_1\di S_2}$ can be identified with a pair,
$(s_1,s_2)$, of points $s_1\in\Si_{\s,S_1}$ and
$s_2\in\Si_{\s,S_2}$. This is possible since the cartesian product
$\Si_{\s,S_1}\times\Si_{\s,S_2}$ is a product in the categorial
sense and hence has projections $\Si_{\s,S_1}\leftarrow
\Si_{\s,S_1}\times\Si_{\s,S_2}\map\Si_{\s,S_2}$. Then the
translation representation of functions is constructed in a
straightforward manner. Thus, let
\begin{eqnarray}
\s(p):\Si_{\s,S_1}\times\Si_{\s,S_2}  &  \map&\Si_{\s,S_1}\nonumber\\
        (s_1,s_2)  &  \mapsto& s_1
\end{eqnarray}
be the canonical projection. Then, if $A_{\s,S_1}\in
C(\Si_{\s,S_1},\mathR)$, the function
\begin{equation}
A_{\s,S_1}\circ\s(p)\in
C(\Si_{\s,S_{1}}\times\Si_{\s,S_{2}},\mathR)
\end{equation}
is such that, for all
$(s_1,s_2)\in\Si_{\s,S_1}\times\Si_{\s,S_2}$,
\begin{equation}
A_{\s,S_1}\circ\s(p)(s_{1},s_{2})=A_{\s,S_1}(s_1).
\end{equation}
Thus we can define
\begin{equation}
        \s(\L{p})(A_{\s,S_1}):=A_{\s,S_1}\circ\s(p).
\end{equation}
Clearly, $\s(\L{p})(A_{\s,S_1})$ can be seen as the representation
of the function symbol $A\di 1\in\F{S_1\di S_2}$.

\section{Theories of Physics in a General Topos}
\label{Sec:ToposAxioms}
\subsection{The Pull-Back Operations}
\subsubsection{The Pull-Back of Physical Quantities.}
Motivated by the above, let us try now to see what can be said
about the scheme in general. Basically, what is involved is the
topos representation of translations of languages. To be more
precise, let $j:S_1\map S$ be an arrow in $\Sys$, so that there is
a translation $\L{j}:\L{S}\map \L{S_1}$ defined by the translation
function $\L{j}:\F{S}\map \F{S_1}$. Now suppose that the systems
$S$ and $S_1$ are represented in the topoi $\tau_\phi(S)$ and
$\tau_\phi(S_1)$ respectively. Then, in these representations, the
function symbols of signature $\Si\map\cal R$  in $\L{S}$ and
$\L{S_1}$ are represented by elements of
$\Hom{\tau_\phi(S)}{\Si_{\phi,S}}{\R_{\phi,S}}$ and
$\Hom{\tau_\phi(S_1)}{\Si_{\phi,S_1}}{\R_{\phi,S_1}}$
respectively.

Our task is to find a  function
\begin{equation}
 \phi(\L{j}):\Hom{\tau_\phi(S)}{\Si_{\phi,S}}
 {\R_{\phi,S}}\map\Hom{\tau_\phi(S_1)}{\Si_{\phi,S_1}}{\R_{\phi,S_1}}
\end{equation}
that can be construed as the topos representation of the
translation $\L{j}:\L{S}\map \L{S_1}$, and hence of the  arrow
$j:S_1\map S$ in $\Sys$. We are particularly interested in seeing
if $\phi(\L{j})$ can be chosen so that the following diagram, (see
\eq{ComDLphi0}) commutes:
\begin{equation}                                \label{ComDLphi}
        \setsqparms[1`1`1`1;1200`700]
        \square[\F{S}`\Hom{\tau_\phi(S)}{\Si_{\phi,S}}
        {\R_{\phi,S}}`\F{S_1}`\Hom{\tau_\phi(S_1)}
        {\Si_{\phi,S_1}}{\R_{\phi,S_1}};
        \rho_{\phi,S}`\L{j}`\phi(\L{j})`\rho_{\phi,S_1}]
\end{equation}
However, as has been emphasised already, it is not clear that one
\emph{should} expect to find a function $\phi(\L{j}):
\Hom{\tau_\phi(S)}{\Si_{\phi,S}}{\R_{\phi,S}} \map
\Hom{\tau_\phi(S_1)}{\Si_{\phi,S_1}}{\R_{\phi,S_1}}$  with this
property. The existence and/or properties of such a function will
be dependent on the theory-type, and it seems unlikely that much
can be said in general about the diagram \eq{ComDLphi}.
Nevertheless, let us see how far we \emph{can} get in discussing
the existence of such a function in general.

Thus, if $\mu\in\Hom{\tau_\phi(S)}{\Si_{\phi,S}}{\R_{\phi,S}}$,
the critical question is if there is some `natural' way whereby
this arrow can be `pulled-back' to give an element
$\phi(\L{j})(\mu)\in\Hom{\tau_\phi(S_1)}
{\Si_{\phi,S_1}}{\R_{\phi,S_1}}$.

The first pertinent remark is that $\mu$ is an arrow in the topos
$\tau_\phi(S)$, whereas the sought-for pull-back will be an arrow
in the topos $\tau_\phi(S_1)$, and so we need a mechanism for
getting from one topos to the other (this problem, of course, does
not arise in classical physics since the topos of every
representation is always $\Set$).

The obvious way of implementing this change of topos is via some
\emph{functor}, $\tau_\phi(j)$ from $\tau_\phi(S)$\ to
$\tau_\phi(S_1)$. Indeed, given such  a functor, an arrow
$\mu:\Si_{\phi,S}\map\R_{\phi,S}$ in $\tau_\phi(S)$ is transformed
to the arrow
\begin{equation}
\tau_\phi(j)(\mu):
\tau_\phi(j)(\Si_{\phi,S})\map\tau_\phi(j)(\R_{\phi,S})
                                \label{taut(A)}
\end{equation}
 in $\tau_\phi(S_1)$.

To convert this to  an arrow from $\Si_{\phi,S_1}$ to
$\R_{\phi,S_1}$, we need to supplement \eq{taut(A)} with a pair of
arrows $\phi(j),\beta_\phi(j)$ in $\tau_\phi(S_1)$ to get the
diagram:
\begin{equation}                                \label{PBPhimu}
        \setsqparms[1`-1`1`0;1000`450]
        \square[\tau_\phi(j)(\Si_{\phi,S})`\tau_\phi(j)(\R_{\phi,S})
                `\Si_{\phi,S_1}`\R_{\phi,S_1};
        \tau_\phi(j)(\mu)`\phi(j)`\beta_\phi(j)`]
\end{equation}
The pull-back, $\phi(\L{j})(\mu)\in
\Hom{\tau_\phi(S_1)}{\Si_{\phi,S_1}}{\R_{\phi,S_1}}$, with respect
to these choices can then be defined as
\begin{equation}
  \phi(\L{j})(\mu):=\beta_\phi(j)\circ \tau_\phi(j)(\mu)\circ\phi(j).
                \label{Def:phi(t)(A)}
\end{equation}
It follows that a key part of the construction of a topos
representation, $\phi$, of $\Sys$ will be to specify the functor
$\tau_\phi(j)$ from $\tau_\phi(S)$ to $\tau_\phi(S_1)$, and the
arrows $\phi(j):\Si_{\phi,S_1}\map\tau_\phi(j)(\Si_{\phi,S})$ and
$\beta_\phi(j):\tau_\phi(j)(\R_{\phi,S})\map\R_{\phi,S_1}$ in the
topos $\tau_\phi(S_1)$. These need to be defined in such a way as
to be consistent with a chain of arrows $S_2\map S_1\map S$.

When applied to the representative $A_{\phi,S}:\Si_{\phi,S}\map
\R_{\phi,S}$ of a physical quantity $A\in\F{S}$, the diagram
\eq{PBPhimu}  becomes (augmented with the upper half)
\begin{equation}                                \label{PBPhiA}
        \xext=3000
        \yext=3000
        \setsqparms[1`-1`1`1;1200`450]
        \square[\tau_\phi(j)(\Si_{\phi,S})`\tau_\phi(j)
        (\R_{\phi,S})`\Si_{\phi,S_1}`\R_{\phi,S_1};
        `\phi(j)`\beta_\phi(j)`\phi(\L{j})(A_{\phi,S})]
        \setsqparms[1`1`1`0;1200`450]
        \putsquare(-1470,450)[\Si_{\phi,S}`\R_{\phi,S}``;
        A_{\phi,S}`\tau_\phi(j)`\tau_\phi(j)`\tau_\phi(j)(A_{\phi,S})]
\end{equation}
The commutativity of \eq{ComDLphi} would then require
\begin{equation}
        \phi(\L{j})(A_{\phi,S})=(\L{j}A)_{\phi,S_1}\label{ClassCom}
\end{equation}
or, in a more expanded notation,
 \begin{equation}
        \phi(\L{j})\circ\rho_{\phi,S}=\rho_{\phi,S_1}\circ\L{j},
                                                                \label{ClassComEx}
\end{equation}
where both the left hand side and the right hand side of
\eq{ClassComEx} are mappings from $\F{S}$ to $\Hom{\tau_\phi(S_1)}
{\Si_{\phi,S_1}}{\R_{\phi,S_1}}$.

Note that the analogous diagram in classical physics is simply
\begin{equation}
        \setsqparms[1`-1`1`1;1000`450]
        \square[\Si_{\s,S}`\mathR`\Si_{\s,S_1}`\mathR;
        A_{\s,S}`\s(j)`\id`\s(\L{j})(A_{\s,S})]
\end{equation}
and the commutativity/pull-back condition \eq{ClassCom} becomes
\begin{equation}
\s(\L{j})(A_{\s,S})=(\L{j}A)_{\phi,S_1}
\end{equation}
which is satisfied by virtue of \eq{ClassPB}.

It is clear from the above that the arrow
$\phi(j):\Si_{\phi,S_1}\map\tau_\phi(j)(\Si_{\phi,S})$ can be
viewed as the topos analogue of the map
$\s(j):\Si_{\s,S_1}\map\Si_{\s,S}$ that arises in classical
physics whenever there is an arrow $j:S_1\map S$.

\subsubsection{The Pull-Back of Propositions.}
More insight can be gained into the nature of the triple
$\la\tau_\phi(j),\phi(j),\beta_\phi(j)\ra$ by considering the
analogous  operation for propositions. First, consider  an arrow
$j:S_1\map S$ in $\Sys$ in classical physics. Associated with this
there is (i) a translation $\L{j}:\L{S}\map \L{S_1}$; (ii) an
associated translation mapping $\L{j}:\F{S}\map \F{S_1}$; and
(iii) a symplectic function $\s(j):\Si_{\s,S_1}\map \Si_{\s,S}$.

Let $K$ be a (Borel) subset of the state space, $\Si_{\s,S}$;
hence $K$ represents a proposition about the system $S$. Then
$\s(j)^*(K):=\s(j)^{-1}(K)$ is a subset of $\Si_{\s,S_1}$ and, as
such, represents a proposition about the system $S_1$.  We say
that $\s(j)^*(K)$ is the \emph{pull-back} to $\Si_{\s,S_1}$ of the
$S$-proposition represented by $K$. The existence of such
pull-backs is part of the consistency of the representation of
propositions in classical mechanics, and it is important to
understand what the analogue of this is in our topos scheme.

Consider the general case with the two systems $S_1,S$ as above.
Then let $K$ be a proposition, represented as a sub-object of
$\Si_{\phi,S}$, with a monic arrow $i_K:
K\hookrightarrow\Si_{\phi,S}$. The question now is if the triple
$\la\tau_\phi(j),\phi(j),\beta_\phi(j)\ra$ can be used to pull $K$
back to give a proposition in $\tau(S_1)$, \ie\ a sub-object of
$\Si_{\phi, S_1}$?

The first requirement is that the functor
$\tau_\phi(j):\tau_{\phi}(S)\map \tau_\phi(S_1)$ should
\emph{preserve monics}; for example by being left-exact. In this
case, the monic arrow $i_K: K\hookrightarrow\Si_{\phi,S}$ in
$\tau_\phi(S)$ is transformed to the monic arrow
\begin{equation}
        \tau_\phi(j)(i_K):\tau_\phi(j)(K)
        \hookrightarrow\tau_\phi(j)(\Si_{\phi,S})
\end{equation}
in $\tau_\phi(S_1)$; thus $\tau_\phi(j)(K)$ is a sub-object of
$\tau_\phi(j)(\Si_{\phi,S})$ in $\tau_\phi(S_1)$. It is a property
of a topos that the pull-back of a monic arrow is monic ; \ie\ if
$M\hookrightarrow Y$ is monic, and if $\psi:X\map Y$, then
$\psi^{-1}(M)$ is a sub-object of $X$.  Therefore, in the case of
interest, the monic arrow $
\tau_\phi(j)(i_K):\tau_\phi(j)(K)\hookrightarrow
\tau_\phi(j)(\Si_{\phi,S})$ can be pulled back along
$\phi(j):\Si_{\phi,S_{1}}\map \tau_\phi(j)(\Si_{\phi,S})$ (see
diagram \eq{PBPhiA}) to give the monic
$\phi(j)^{-1}(\tau_\phi(j)(K))\subseteq\Si_{\phi,S_1}$. This is a
candidate for the pull-back of the proposition represented by the
sub-object $K\subseteq \Si_{\phi,S}$.

In conclusion, propositions can be pulled-back provided that the
functor $\tau_\phi(j):\tau_{\phi}(S)\map \tau_\phi(S_1)$ preserves
monics; for example, by being left-exact. In fact, the property of
being left-exact is so natural that we shall add it to our list of
postulates; see below.

\subsection{The Topos Rules for Theories of Physics}
\label{SubSec:GeneralToposAxioms} We will now present  our general
rules for using topos theory in the mathematical representation of
physical systems and their theories. {\definition The category
${\cal M}(\Sys)$ is defined as follows:
\begin{enumerate}
\item The objects of ${\cal M}(\Sys)$ are the topoi that are to be
used in representing the systems in $\Sys$.

\item The arrows from $\tau_1$ to $\tau_2$ are defined to be
the  left-exact functors from $\tau_1$ to $\tau_2$.
\end{enumerate}
}

{\definition The  rules for using topos theory are as follows:
}\label{D_GeneralToposRules}
\begin{enumerate}
\item A \emph{topos realisation}, $\phi$, of $\Sys$ in
${\cal M}(\Sys)$ is an assignment, to each system $S$ in $\Sys$,
of a triple $\phi(S)=\la\rho_{\phi, S},\L{S},\tau_\phi(S)\ra$
where:
\begin{enumerate}
        \item $\tau_\phi(S)$ is the topos in  ${\cal M}(\Sys)$ in
         which the theory-type applied to system $S$ is to be realised.

       \item $\L{S}$ is the local language that is associated with $S$.
        This is independent of the realisation, $\phi$, of $\Sys$
        in ${\cal M}(\Sys)$.

\item $\rho_{\phi,S}:\L{S}\sq\tau_\phi(S)$ is
       a representation of the local language $\L{S}$ in the
       topos $\tau_\phi(S)$.

\item In addition, for each arrow $j:S_1\map S$ in $\Sys$ there
is a triple  $\la\tau_\phi(j)$,$\phi(j)$, $\beta_\phi(j)\ra$ that
interpolates between $\rho_{\phi,S}:\L{S}\sq\tau_\phi(S)$ and
$\rho_{\phi,S_1}:\L{S_1}\sq\tau_\phi(S_1)$; for details see below.
\end{enumerate}

\item
\begin{enumerate}
\item The representations, $\rho_{\phi,S}(\Si)$ and
$\rho_{\phi,S}(\R)$,  of the ground symbols $\Si$ and $\R$ in
$\L{S}$ are denoted $\Si_{\phi,S}$ and $\R_{\phi, S}$,
respectively. They are known as the `state object' and
`quantity-value object' in $\tau_\phi(S)$.

\item The representation by $\rho_{\phi,S}$ of each function symbol
$A:\Si\map\R$ of the system $S$ is an arrow, $\rho_{\phi,S}(A):
\Si_{\phi,S}\map\R_{\phi,S}$ in $\tau_\phi(S)$; we will usually
denote this arrow as $A_{\phi, S}:\Si_{\phi,S}\map\R_{\phi,S}$.

\item Propositions about the system $S$ are represented by sub-objects
of $\Si_{\phi,S}$. These will typically be of the form
$A_{\phi,S}^{-1}(\Xi)$, where $\Xi$ is a sub-object of
$\R_{\phi,S}$.\footnote{Here, $A_{\phi,S}^{-1}(\Xi)$ denotes the
sub-object of $\Si_{\phi,S}$ whose characteristic arrow is
$\chi_{\Xi}\circ A_{\phi,S}:\Si_{\phi,S}\map
\Omega_{\tau_\phi(S)}$, where $\chi_{\Xi}:\R_{\phi,S}\map
\Omega_{\tau_\phi(S)}$ is the characteristic arrow of the
sub-object $\Xi$.}
\end{enumerate}

\item Generally, there are  no `microstates' for the system $S$;
\ie\  no global elements (arrows $1\map \Si_{\phi,S}$) of the
state object $\Si_{\phi,S}$; or, if there are any, they may not be
enough to determine $\Si_{\phi,S}$ as an object in $\tau_\phi(S)$.

Instead, the role of a state is played by a `truth sub-object'
$\TO$ of $P\Si_{\phi,S}$.\footnote{In classical physics, the truth
object corresponding to a microstate $s$ is the collection of all
propositions that are true in the state $s$.} If
$\ga\in\Sub{\Si_{\phi,S}}\simeq\Ga(P\Si_{\phi,S})$, the `truth of
the proposition represented by $\ga$' is defined to be
\begin{equation}
\TValM{\name\ga\in\TO}=
        \Val{\tilde\ga\in\tilde\TO}_\phi\, \langle\ga,\TO\rangle
\end{equation}
See paper II for full information on the idea of a `truth object'
\cite{DI(2)}.

\item  There is a `unit object' $1_{{\cal M}(\Sys)}$ in ${\cal
M}(\Sys)$ such that if $1_\Sys$ denotes the trivial system in
$\Sys$ then, for all topos realisations $\phi$,
    \begin{equation}
        \tau_\phi(1_\Sys)=1_{{\cal M}(\Sys)}.
    \end{equation}

Motivated by the results for quantum theory (see Section
\ref{SubSubSec:TransSum}), we postulate that the unit object
$1_{{\cal M}(\Sys)}$ in ${\cal M}(\Sys)$ is  the category of sets:
    \begin{equation}
            1_{{\cal M}(\Sys)}=\Set.
    \end{equation}

\item To each arrow $j:S_1\map S$ in $\Sys$, we have the following:
\begin{enumerate}
\item There is a translation $\L{j}:\L{S}\map\L{S_1}$.
This is specified by a map between function symbols:
$\L{j}:\F{S}\map\F{S_1}$.

\item With the translation
$\L{j}:\F{S}\map\F{S_1}$ there is associated a corresponding
function
\begin{equation}
        \phi(\L{j}):\Hom{\tau_\phi(S)}{\Si_{\phi,S}}{\R_{\phi,S}}
        \map\Hom{\tau_\phi(S_1)}{\Si_{\phi,S_1}}{\R_{\phi,S_1}}.
\end{equation}
These may, or may not, fit together in the commutative diagram:
\begin{equation}           \label{Diag_CommOfTranslationAndRep}
        \setsqparms[1`1`1`1;1200`700]
        \square[\F{S}`\Hom{\tau_\phi(S)}{\Si_{\phi,S}}
        {\R_{\phi,S}}`\F{S_1}`\Hom{\tau_\phi(S_1)}
        {\Si_{\phi,S_1}}{\R_{\phi,S_1}};
        \rho_{\phi,S}`\L{j}`\phi(\L{j})`\rho_{\phi,S_1}]
\end{equation}
\item The function  $\phi(\L{j}): \Hom{\tau_\phi(S)}{\Si_{\phi,S}}
{\R_{\phi,S}} \map\Hom{\tau_\phi(S_1)}{\Si_{\phi,S_1}}
{\R_{\phi,S_1}}$ is built from the following ingredients. For each
topos realisation $\phi$, there is a triple
$\la\tau_\phi(j),\phi(j),\beta_\phi(j)\ra$ where:
\begin{enumerate}
\item $\tau_\phi(j):\tau_\phi(S)\map\tau_\phi(S_1)$ is a left-exact
functor; \ie\ an arrow in ${\cal M}(\Sys)$.

\item
$\phi(j):\Si_{\phi,S_1}\map \tau_\phi(j)\big(\Si_{\phi,S}\big)$ is
an arrow in $\tau_\phi(S_1)$.

\item
$\beta_\phi(j):\tau_\phi(j)\big(\R_{\phi,S}\big)
\map\R_{\phi,S_1}$ is an arrow in $\tau_\phi(S_{1})$.
\end{enumerate}
These fit together in the diagram
\begin{equation}                                \label{PBPhiA2}
        \xext=3000
        \yext=3000
        \setsqparms[1`-1`1`1;1200`450]
        \square[\tau_\phi(j)(\Si_{\phi,S})`\tau_\phi(j)(\R_{\phi,S})
        `\Si_{\phi,S_1}`\R_{\phi,S_1};
        `\phi(j)`\beta_\phi(j)`\phi(\L{j})(A_{\phi,S})]
        \setsqparms[1`1`1`0;1200`450]
        \putsquare(-1470,450)[\Si_{\phi,S}`\R_{\phi,S}``;
        A_{\phi,S}`\tau_\phi(j)`\tau_\phi(j)`\tau_\phi(j)(A_{\phi,S})]
\end{equation}
The arrows $\phi(j)$ and $\beta_\phi(j)$ should behave
appropriately under composition of arrows in $\Sys$.

The commutativity of the diagram \eq{Diag_CommOfTranslationAndRep}
is equivalent to the relation
\begin{equation}
\phi(\L{j})(A_{\phi,S})=[\L{j}(A)]_{\phi,S_1}\label{CDpullback}
\end{equation}
for all $A\in\F{\phi,S}$. As we keep emphasising, the satisfaction
or otherwise of this relation will depend on the theory-type and,
possibly, the representation $\phi$.

\item If a proposition in $\tau_\phi(S)$ is represented by the
monic arrow, $K\hookrightarrow\Si_{\phi,S}$, the `pull-back' of
this proposition to $\tau_\phi(S_1)$ is defined to be
$\phi(j)^{-1}\big(\tau_\phi(j)(K)\big)\subseteq\Si_{\phi,S_1}$.
\end{enumerate}

\item
\begin{enumerate}
\item If $S_1$ is a sub-system of $S$, with an associated arrow
$i:S_1\map S$ in $\Sys$ then, in the diagram in \eq{PBPhiA2}, the
arrow $\phi(j):\Si_{\phi,S_1}\map\tau_\phi(j)(\Si_{\phi,S})$ is a
monic arrow in $\tau_\phi(S_{1})$.

In other words, $\Si_{\phi,S_1}$ is a sub-object of
$\tau_\phi(j)(\Si_{\phi,S})$, which is denoted
\begin{equation}
\Si_{\phi,S_1}\subseteq
        \tau_\phi(j)(\Si_{\phi,S}).  \label{SubSysSS}
\end{equation}

We may also want to conjecture
\begin{equation}
\R_{\phi,S_1}\simeq
            \tau_\phi(j)\big(\R_{\phi,S}\big).  \label{RsimR}
\end{equation}

\item Another possible conjecture is the following: if $j:S_1\map S$
is an epic arrow in $\Sys$, then, in the diagram in \eq{PBPhiA2},
the arrow $\phi(j):\Si_{\phi,S_1}\map \tau_\phi(j)(\Si_{\phi,S})$
is an epic arrow in $\tau_\phi(S_1)$.

In particular, for the epic arrow $p_1:S_1\di S_2\map S_1$, the
arrow $\phi(p_1):\Si_{\phi,S_1\di S_2}\map
\tau_\phi\big(\Si_{\phi,S_1}\big)$ is an epic arrow in the topos
$\tau_\phi(S_1\di S_2)$.
\end{enumerate}
\end{enumerate}

One should not read Rule 2.\ above as implying that the choice of
the state object and quantity-value object are \emph{unique} for
any give system $S$. These objects would at best be selected only
up to isomorphism in the topos $\tau(S) $. Such morphisms in the
topos $\tau(S)$\footnote{Care is needed not to confuse morphisms
in the topos $\tau(S)$ with morphisms in the category ${\cal
M}(\Sys)$ of topoi. An arrow from the object $\tau(S)$ to itself
in the category ${\cal M}(\Sys)$ is a left-exact morphism in the
topos $\tau(S)$. However, not every arrow in $\tau(S)$ need arise
in this way, and an important role can be expected to be played by
arrows of this type. A good example is when $\tau(S)$ is the
category of sets, $\Set$. Typically, $\tau_\phi(j):\Set\map\Set$
is the identity, but there are many morphisms from an object $O$
in $\Set$ to itself: they are just the functions from $O$ to $O$.}
can be expected to play a key role  in developing the topos
analogue of the important idea of a \emph{symmetry}, or
\emph{covariance} transformation of the theory. These ideas were
discussed briefly in Paper III \cite{DI(3)}.

In the example of classical physics,  for all systems we have
$\tau(S)=\Set$ and $\Si_{\s,S}$ is a symplectic manifold, and the
collection of all symplectic manifolds is a category. It would be
elegant if we could assert that, in general, for a given
theory-type the possible state objects in a given topos $\tau$
form the objects of an \emph{internal} category in $\tau$.
However, to make such a statement would require a general theory
of state objects and, at the moment, we do not have such a thing.

From a more conceptual viewpoint we note that the `similarity' of
our axioms to those of standard classical physics is reflected in
the fact that (i) physical quantities are represented by arrows
$A_{\phi,S}:\Si_{\phi,S}\map\R_{\phi,S}$; (ii) propositions are
represented by sub-objects of $\Si_{\phi,S}$; and (iii)
propositions are assigned truth values. Thus any theory satisfying
these axioms `looks' like classical physics, and has an associated
neo-realist interpretation.

\section{The General Scheme applied to Quantum Theory}
\label{Sec:ReviewQT}
\subsection{Background Remarks}
We now want to study the extent to which our `rules' apply to the
topos representation of quantum theory.

For a quantum system with (separable) Hilbert space $\Hi$, the
appropriate topos (what we earlier called $\tau_\phi(S)$) is
$\SetH{}$: the category of presheaves over the category (actually,
partially-ordered set) $\V{}$ of unital, abelian von Neumann
subalgebras of the algebra, $\BH$,\ of bounded operators on $\cal
H$.

From a physical perspective we can think of the objects in
$\V{}$---\ie\ the commutative subalgebras of $\BH$---as the
`contexts' (or `world views', or `windows on reality') with
respect to which our generalised truth values of propositions are
assigned. In the normal, instrumentalist interpretation of quantum
theory, a context is therefore a collection of physical variables
that can be measured simultaneously. The physical significance of
this contextual logic is discussed at length in \cite{IB98, IB99,
IB00, IB02, IB00b} and \cite{DI(2),DI(3)}.

A particularly important object in $\SetH{}$ is the \emph{spectral
presheaf} $\Sig$, where, for each $V$, $\Sig_V$ is defined to be
the Gel'fand spectrum of the abelian algebra $V$. The sub-objects
of $\Sig$ can be identified as the topos representations of
propositions, just as the subsets of $\S$ represent propositions
in classical physics.

In \cite{DI(3)}, several closely related choices for a
quantity-value object $\R_\phi$ in $\SetH{}$ were discussed. We
concentrate here on the presheaf $\SR$ of real-valued,
order-reversing functions. Physical quantities
$A:\Si\rightarrow\R$, which correspond to self-adjoint operators
$\hat A$, are represented by natural transformations/arrows
$\dasBo{A}:\Sig\rightarrow\SR$.  The mapping $\hat
A\mapsto\dasBo{A}$ is injective. For brevity, we write
$\dasB{A}:=\dasBo{A}$.\footnote{Note that this is \emph{not} the
same as the convention used in paper III \cite{DI(3)}, where
$\dasB{A}$ denoted a different natural transformation.} The
arguments given in this section apply in similar form to the other
possible choices for the quantity-value object.

\paragraph{Geometric Morphisms.}
Our constructions require a left-exact functor between two topoi,
and one of the natural sources of such things is a \emph{geometric
morphism}. This fundamental concept in topos theory is defined as
follows \cite{MM92}.
\begin{definition}
A \emph{geometric morphism} $\phi:{\cal F}\map{\cal E}$ between
topoi $\cal F$ and $\cal E$ is a pair of functors $\phi^*:{\cal
E}\map{\cal F}$ and $\phi_*:{\cal F}\map \cal E$ such that
\begin{enumerate}
    \item[(i)] $\phi^*\dashv \phi_*$, \ie\ $\phi^*$ is left adjoint
    to $\phi_*$;
    \item[(ii)] $\phi^*$ is left exact, \ie\ it preserves
    all finite limits.
\end{enumerate}
\end{definition}

Geometric morphisms are very important because they are the topos
equivalent of continuous functions. More precisely, if $X$ and $Y$
are topological spaces, then any continuous function $f:X\map Y$
induces a geometric morphism between the topoi ${\rm Sh}(X)$ and
${\rm Sh}(Y)$ of sheaves on $X$ and $Y$ respectively.  In fact,
just as the arrows in the category of topological spaces are
continuous functions, so in any category whose objects are topoi,
the arrows are normally defined to be geometric morphisms.

The key result for us is the following theorem (\cite{MM92} p359):
\begin{theorem}\label{Th:phiC->D}
If $\varphi:{\cal C}\map {\cal D}$ is a functor between categories
$\cal C$ and $\cal D$, then it induces a geometric morphism (also
denoted $\varphi$)
\begin{equation}
        \varphi:\SetC{{\cal C}}\map\SetC{{\cal D}}
\end{equation}
for which the functor $\varphi^*:\SetC{{\cal D}}\map\SetC{{\cal
C}}$ takes a functor $\ps{F}:{\cal D}\map\Set$ to the functor
\begin{equation}
   \varphi^*(\ps{F}):=\ps{F}\circ\varphi^\op    \label{Def:phi*(F)}
\end{equation}
from $\cal C$ to $\Set$.

In addition, $\varphi^*$ has a left adjoint $\varphi_!$; \ie\
$\varphi_!\dashv\varphi^*$.
\end{theorem}
The morphism $\varphi^{*}:\SetC{{\cal D}}\map\SetC{{\cal C}}$ is
called the \emph{inverse image part} of the geometric morphism
$\varphi$; the morphism $\varphi_*:\SetC{{\cal C}}\map\SetC{{\cal
D}}$ is called the \emph{direct image part}.

We will use this important theorem in several crucial places.

\subsection{The Translation Representation for  a Disjoint
Sum of Quantum Systems}\label{SubSubSec:TransSum} Let $\Sys$ be a
category whose objects are systems that can be treated using
quantum theory. Let $\L{S}$ be the local language of a system $S$
in $\Sys$ whose quantum Hilbert space is denoted $\Hi_S$. We
assume that to each function symbol, $A:\Si\map\R$, in $\L{S}$
there is associated a self-adjoint operator $\hat
A\in\mathcal{B(H} _{S}),$\footnote{More specifically, one could
postulate that the elements of $\F{S}$ are associated with
self-adjoint operators in some unital von Neumann subalgebra of
$\mathcal{B(H}_{S})$.} and that the map
\begin{eqnarray}
         \F{S} &\rightarrow& \BH_\sa\\
        A &\mapsto& \hat A
\end{eqnarray}
is injective (but not necessarily surjective, as we will see in
the case of a disjoint sum of quantum systems).

We consider first arrows of the form
\begin{equation}
S_{1}\overset{i_{1}}{\rightarrow}S_{1}\sqcup S_{2}\overset{i_{2}
}{\leftarrow}S_{2}
\end{equation}
from the components $S_{1}$, $S_{2}$ to a disjoint sum
$S_{1}\sqcup S_{2}$; for convenience we write $i:=i_{1}$.  The
systems $S_{1}$, $S_{2}$ and $S_{1}\sqcup S_{2}$ have the Hilbert
spaces $\Hi_1$, $\Hi_2$ and $\Hi_1\oplus\Hi_2$, respectively.

As always, the translation $\L{i}$ goes in the opposite direction
to the arrow $i$, so
\begin{equation}
\L{i}:\F{S_1\sqcup S_2}\map \F{S_1}.
\end{equation}
Then our first step is find an  `operator translation' from the
relevant self-adjoint operators in $\Hi_1\oplus\Hi_2$ to those in
$\Hi_1$,

To do this, let $A$ be a function symbol in $\F{S_1\sqcup S_2}$.
In Section \ref{SubSubSec:ATDS}, we argued that $\F{S_1\sqcup
S_2}\simeq\F{S_1}\times\F{S_2}$ (as in \eq{FS1sumS2}), and hence
we introduce the notation $A=\la A_1, A_2\ra$, where
$A_1\in\F{S_1}$ and $A_2\in\F{S_2}$. It is then natural to assume
that the quantisation scheme is such that the operator, $\hat A$,
on $\Hi_1\oplus\Hi_2$ can be decomposed as $\hat A=\hat
A_1\oplus\hat A_2$, where the operators $\hat A_1$ and $\hat A_2$
are defined on $\Hi_1$ and $\Hi_2$ respectively, and correspond to
the function symbols $A_1$ and $A_2$.\footnote{It should be noted
that our scheme does not use all the self-adjoint operators on the
direct sum $\Hi_1\oplus\Hi_2$: only the `block diagonal' operators
of the form $\hat A=\hat A_1\oplus\hat A_2$ arise.} Then the
obvious operator translation is $\hat A\mapsto\hat
A_1\in\mathcal{B(H}_{1})_{sa}$.

We now consider the general rules in the  Definition
\ref{D_GeneralToposRules} and see to what extent they apply in the
example of quantum theory.

\textbf{1.} As we have stated several times, the topos
$\tau_{\phi}(S)$ associated with a quantum system $S$ is
\begin{equation}
\tau_{\phi}(S)=\Set^{\mathcal{V(H}_{S})^{op}}.
\end{equation}
Thus (i) the objects of the category $\mathcal{M}(\Sys)$ are topoi
of the form $\Set^{\mathcal{V(H}_{S})^{op}}$, $S\in\Ob\Sys$; and
(ii) the arrows between two topoi are defined to be  left-exact
functors. In particular, to each arrow $j:S_1\map S$ in $\Sys$
there must correspond a left-exact functor
$\tau_\phi(j):\tau_\phi(S)\map\tau_\phi(S_{1})$. Of course, the
existence of these functors in the quantum case has yet to be
shown.

\textbf{2.} The realisation $\rho_{\phi,S}:\L{S}\rightsquigarrow
\tau_\phi(S)$ of the language $\L{S}$ in the topos $\tau_\phi(S)$
is given as follows. First, we define the state object
$\Si_{\phi,S}$ to be  the spectral presheaf,
$\Sig^{\mathcal{V(H}_{S})}$, over $\mathcal{V(H}_{S}\mathcal{)}$,
the context category of $\mathcal{B(H}_{S})$. To keep the notation
brief, we will denote\footnote{Presheaves are always denoted by
symbols that are underlined.}
 $\Sig^{\mathcal{V(H}_{S})}$ as $\Sig^{\Hi_{S}}$.

Furthermore, we define the quantity-value object, $\R_{\phi,S}$,
to be the presheaf $\SR{}^{\;\Hi_{S}}$ that was defined in paper
III \cite{DI(3)}. Finally, we define
\begin{equation}
A_{\phi,S}:=\dasB{A},
\end{equation}
for all $A\in\F{S}$. Here $\dasB{A}:\Sig^{\Hi_{S}
}\rightarrow\SR{}^{\Hi_S}$ is constructed using the Gel'fand
transforms of the (outer)\ daseinisation of $\hat A$, for details
see paper III.

\textbf{3.} The truth object $\mathbb{T}^{\psi}$ corresponding to
a pure state $\psi$ was discussed in paper II \cite{DI(2)}.

\textbf{4.} Let $\Hi=\mathC$ be the one-dimensional Hilbert space,
corresponding to the trivial quantum system $1$. There is exactly
one abelian subalgebra of $\mathcal{B}(\mathC)\simeq\mathC$,
namely $\mathC$ itself. This leads to
\begin{equation}
\tau_{\phi}(1_{\Sys})=\Set^{\{*\}}\simeq\Set=1_{\mathcal{M}(\Sys)}.
\end{equation}

\textbf{5.} Let $A\in \F{S_{1}\sqcup S_{2}}$ be a function symbol
for the system $S_1\sqcup S_2$. Then, as discussed above, $A$ is
of the form $A=\la A_1, A_2\ra$ (compare equation \eq{FS1sumS2}),
which corresponds to a self-adjoint operator $\hat
A_1\oplus\hat{A}_2\in\mathcal{B(H}_{1}\oplus\Hi_2)_{\sa}$. The
topos representation of $A$ is the natural transformation
$\breve\delta(\la A_1, A_2\ra):\Sig^{\Hi_1\oplus\Hi_2}
\map\SR{}^{\;\Hi_1\oplus\Hi_2}$, which is defined at each stage
$V\in\Ob{\mathcal{V(H}_{1}\oplus\Hi_2)}$ as
\begin{eqnarray}
\breve\delta(\la A_1,A_2\ra)_V:\Sig^{\Hi_1\oplus\Hi_2}_V
  &\map&\SR{}^{\;\Hi_1\oplus\Hi_2}_V \nonumber  \\
\l &\mapsto&\{V^\prime\mapsto\l|_{V^{\prime}}(\delta(
\hat{A_1}\oplus\hat{A_2})_{V^{\prime}})\mid V^{\prime}\subseteq
V\} \label{Def:dasB(A1+A2)}
\end{eqnarray}
where the right hand side \eq{Def:dasB(A1+A2)} denotes an
order-reversing function.

We will need the following:
\begin{lemma}
\label{L_DaseinisationAndDirectSum}Let
$\hat{A_1}\oplus\hat{A_2}\in\mathcal{B(H}_{1}\oplus\Hi_2)_{\sa}$,
and let $V=V_1\oplus V_2\in\Ob{\mathcal{V}(\Hi_1\oplus\Hi_2)}$
such that $V_1\in\Ob{\V{1}}$ and $V_2\in\Ob{\V{2}}$. Then
\begin{equation}
\delta(\hat{A_1}\oplus\hat{A_2})_{V}=\delta(\hat{A_1})_{V_1}
\oplus\delta(\hat{A_2})_{V_2}.
\label{das(A1+A2)=}
\end{equation}

\end{lemma}

\begin{proof}
Every projection $\hat Q\in V$ is of the form $\hat Q=\hat
Q_{1}\oplus\hat Q_{2}$ for unique projections $\hat
Q_{1}\in\mathcal{P(H}_{1})$ and $\hat Q_{2}
\in\mathcal{P(H}_{2})$. Let $\P\in\PH$ be of the form
$\P=\P_1\oplus\P_2$ such that $\P_1\in\mathcal{P(H}_1)$ and
$\P_2\in\mathcal{P(H}_1)$. The largest projection in $V$ smaller
than or equal to $\P$, \ie\ the inner daseinisation of $\P$ to
$V$, is
\begin{equation}
\dastoi{V}{P}=\hat{Q}_{1}\oplus\hat{Q}_{2},
\end{equation}
where $\hat{Q}_{1}\in\mathcal{P(}V_1)$ is the largest projection
in $V_1$ smaller than or equal to $\P_{1}$, and $\hat{Q}_{2}%
\in\mathcal{P(}V_2)$ is the largest projection in $V_2$ smaller
than or equal to $\P_{2}$, so
\begin{equation}
\dastoi{V}{P}=\delta(\P_{1})_{V_1}\oplus\delta(\P_{2})_{V_2}.
\end{equation}
This implies $\delta(\hat{A}\oplus\hat{B}
)_{V}=\delta(\hat{A})_{V_1}\oplus\delta(\hat{B})_{V_2}$, since
(outer) daseinisation of a self-adjoint operator just means inner
daseinisation of the projections in its spectral family, and all
the projections in the spectral family of $\A\oplus\hat B$ are of
the form $\P=\P_1\oplus\P_2$.
\end{proof}

As discussed in Section \ref{Sec:ToposAxioms}, in order to mimic
the construction that we have in the classical case, we need  to
pull back the arrow/natural transformation $\breve\delta(\la
A_1,A_2\ra): \Sig^{\Hi_1\oplus\Hi_2}
\map\SR{}^{\;\Hi_1\oplus\Hi_2}$  to obtain an arrow from
$\Sig^{\Hi_1}$ to $\SR{}^{\;\Hi_1}$. Since we decided that the
translation on the level of operators sends
$\hat{A_1}\oplus\hat{A_2}$ to $\hat{A_1}$, we expect that this
arrow from $\Sig^{\Hi_1}$ to $\SR{}^{\;\Hi_1}$ is $\dasB{A_1}$. We
will now show how this works.

The presheaves $\Sig^{\Hi_1\oplus\Hi_2}$ and $\Sig^{\Hi_1}$ lie in
different topoi, and in order to `transform' between them we need
we need a (left-exact) functor from the topos
$\Set^{\mathcal{V(\Hi}_1\oplus\Hi_2)^{op}}$to the topos
$\SetH{1}$: this is the functor
$\tau_\phi(j):\tau_\phi(S)\map\tau_\phi(S_1)$ in \eq{PBPhiA2}. One
natural place to look for such a functor is as the inverse-image
part of a geometric morphism from $\Set^{\mathcal{V(H}_1)^{op}}$
to $\Set^{\mathcal{V(H}_{1}\oplus\Hi_2)^{\op}}$. According to
Theorem \ref{Th:phiC->D}, one source of such a geometric morphism,
$\mu$, is a functor
\begin{equation}
        m:\V{1}\map \mathcal{V(H}_{1}\oplus\Hi_2),
\end{equation}
and the obvious choice for this is
\begin{equation}
        m(V):=V\oplus \mathC\hat 1_{\Hi_2}
\end{equation}
for all $V\in\Ob{\V{1}}$. This function from $\Ob{\V{1}}$ to
$\Ob{\mathcal{V(H}_{1}\oplus\Hi_2)}$ is clearly order preserving,
and hence $m$ is a genuine functor.

Let $\mu$ denote the geometric morphism induced by $m$. The
inverse-image functor of $\mu$ is given by
\begin{eqnarray}
\mu^*:\Set^{\mathcal{V(H}_1\oplus\Hi_2)^\op}&\map&
                \Set^{\mathcal{V(H}_{1})^\op}         \\
                \ps{F}  &  \mapsto& \ps{F}\circ m^\op.
\end{eqnarray}
This means that, for all $V\in \Ob{\V{1}}$, we have
\begin{equation}
(\mu^*\ps{F}^{\Hi_1\oplus\Hi_2})_V=\ps{F}^{\Hi_1\oplus\Hi_2}_{m(V)}=
\ps{F}^{\Hi_1\oplus\Hi_2}_{V\oplus \mathC\hat
1_{\Hi_2}}.\label{mu*Si+2}
\end{equation}
For example, for the spectral presheaf we get
\begin{equation}
(\mu^*\Sig^{\Hi_1\oplus\Hi_2})_V=\Sig^{\Hi_1\oplus\Hi_2}_{m(V)}=
        \Sig^{\Hi_1\oplus\Hi_2}_{V\oplus \mathC\hat 1_{\Hi_2}}.
\end{equation}
This is the functor that is denoted
$\tau_\phi(j):\tau_\phi(S_1)\map\tau_\phi(S)$ in  \eq{PBPhiA2}.

We next need to find an arrow
$\phi(i):\Sig^{\Hi_1}\map\mu^*\Sig^{\Hi_1\oplus\Hi_2}$ that is the
analogue of the arrow
$\phi(j):\Si_{\phi,S_1}\map\tau_\phi(j)(\Si_{\phi,S})$ in
\eq{PBPhiA2}.

For each $V$, the set $(\mu^*\Sig^{\Hi_1\oplus\Hi_2})_V=
\Sig^{\Hi_1\oplus\Hi_2}_{V\oplus \mathC\hat 1_{\Hi_2}}$ contains
two types of spectral elements $\l$: the first type are those $\l$
such that $\l(\hat 0_{\Hi_1}\oplus\hat 1_{\Hi_2})=0$. Then,
clearly, there is some $\tilde\l\in\Sig^{\Hi_1}_V$ such that
$\tilde\l(\A)=\l(\A\oplus\hat 0_{\Hi_2})=\l(\A\oplus\hat
1_{\Hi_2})$ for all $\A\in V_\sa$. The second type of spectral
elements $\l\in\Sig^{\Hi_1\oplus\Hi_2}_{V\oplus \mathC\hat
1_{\Hi_2}}$ are such that $\l(\hat 0_{\Hi_1}\oplus\hat
1_{\Hi_2})=1$. In fact, there is exactly one such $\l$, and we
denote it by $\l_0$. This shows that
$\Sig^{\Hi_1\oplus\Hi_2}_{V\oplus \mathC\hat 1_{\Hi_2}}
\simeq\Sig^{\Hi_1}_V\cup\{\l_0\}$. Accordingly, at each stage $V$,
the mapping $\phi(i)$ sends each $\tilde\l\in\Sig^{\Hi_1}_V$ to
the corresponding $\l\in\Sig^{\Hi_1\oplus\Hi_2}_{V\oplus
\mathC\hat 1_{\Hi_2}}$.

The presheaf $\SR{}^{\;\Hi_1\oplus\Hi_2}$ is given at each stage
$W\in\Ob{\mathcal{V}(\Hi_1\oplus\Hi_2)}$ as the order-reversing
functions $\nu:\downarrow\!\!\!W\rightarrow\mathR$, where
$\downarrow\!\!\!W$ denotes the set of unital, abelian von Neumann
subalgebras of $W$. Let $W=V\oplus\mathC\hat 1_{\Hi_2}$. Clearly,
there is a bijection between the sets $\downarrow\!\!
W\subset\Ob{\mathcal{V}(\Hi_1\oplus\Hi_2)}$ and $\downarrow\!\! V
\subset\Ob{\V{}}$. We can thus identify
\begin{equation}
        (\mu^*\SR{}^{\;\Hi_1\oplus\Hi_2})_V=
        \SR{}^{\;\Hi_1\oplus\Hi_2}_{V\oplus \mathC\hat 1_{\Hi_2}}
        \simeq\SR^{\Hi_1}_V
\end{equation}
for all $V\in\Ob{\V{}}$. This gives an isomorphism $\beta_\phi(i):
\mu^*\SR{}^{\;\Hi_1\oplus\Hi_2}\map\SR{}^{\;\Hi_1}$, which
corresponds to the arrow
$\beta_\phi(j):\R_{\phi,S_1}\map\tau_\phi(j)(\R_{\phi,S})$ in
\eq{PBPhiA2}.

Now consider the arrow $\breve\delta(\la A_1, A_2\ra):
\Sig^{\Hi_1\oplus\Hi_2}\map\SR{}^{\;\Hi_1\oplus\Hi_2}$. This is
the analogue of the arrow $A_{\phi,S}:\Si_{\phi,S}\map\R_{\phi,S}$
in \eq{PBPhiA2}. At each stage
$W\in\Ob{\mathcal{V}(\Hi_1\oplus\Hi_2)}$, this arrow is given by
the (outer) daseinisation $\delta(\A_1\oplus\A_2)_{W^{\prime}}$
for all $W^{\prime}\in\downarrow\!\!W$. According to Lemma
\ref{L_DaseinisationAndDirectSum}, we have
\begin{equation}
        \delta(\hat{A}_1\oplus\hat{A}_2)_{V\oplus\mathC\hat 1_{\Hi_2}}=
\dasto{V}{A_1}\oplus\delta(A_2)_{\mathC\hat 1_{\Hi_2}}
        =\dasto{V}{A_1}\oplus\rm{max(sp}(\A_2))\hat 1_{\Hi_2}
\end{equation}
for all $V\oplus\mathC\hat
1_{\Hi_2}\in\Ob{\mathcal{V}(\Hi_1\oplus\Hi_2)}$. This makes clear
how the arrow
\begin{equation}
\mu^*(\breve\delta(\la A_1, A_2\ra)):\mu^*\Sig^{\Hi_1\oplus\Hi_2}
\map\mu^*\SR{}^{\;\Hi_1\oplus\Hi_2}
\end{equation}
is defined. Our conjectured pull-back/translation representation
is
\begin{equation}
        \phi(\L{i})\big(\breve\delta(\la A_1, A_2\ra)\big):=
    \phi(i)\circ\mu^*(\breve\delta(\la A_1, A_2\ra))
    \circ\beta_\phi(i):\Sig^{\Hi_1}\map\SR{}^{\;\Hi_1}.
\end{equation}
Using the definitions of $\phi(i)$ and $\beta_\phi(i)$, it becomes
clear that
\begin{equation}
\phi(i)\circ\mu^*(\breve\delta(\la A_1, A_2\ra))
\circ\beta_\phi(i)=\dasB{A_1}.
\end{equation}
Hence,  the commutativity condition in \eq{CDpullback} is
satisfied for arrows in $\Sys$ of the form $i_{1,2}:S_{1,2}\map
S_1\sqcup S_2$.

\subsection{The Translation Representation for Composite
Quantum Systems}\label{SubSec:TranslCompQT} We now consider arrows
in $\Sys$ of the form
\begin{equation}
S_{1}\overset{p_{1}}{\leftarrow}S_{1}\di S_{2}\overset{p_{2}
}{\rightarrow}S_{1},
\end{equation}
where the quantum systems $S_{1}$, $S_{2}$ and $S_{1}\di S_{2}$
have the Hilbert spaces $\Hi_1$, $\Hi_2$ and $\Hi_1\otimes \Hi_2$,
respectively.\footnote{As usual, the composite system $S_{1}\di
S_{2}$ has as its Hilbert space the tensor product of the Hilbert
spaces of the components.}

The canonical translation\footnote{As discussed in Section
\ref{SubSubSec:ArrTranComp}, this translation, $\L{p_1}$,
transforms a physical quantity $A_1$ of system $S_{1}$ into a
physical quantity $A_1\di1$, which is the `same' physical quantity
but now seen as a part of the composite system $S_{1}\di S_{2}$.
The symbol $1$ is the trivial physical quantity: it is represented
by the operator $\hat{1}_{\Hi_2}$.} $\L{p_1}$ between the
languages $\L{S_1}$ and $\L{S_{1}\di S_{2}}$ (see Section
\ref{SubSubSec:ArrTranComp}) is such that if $A_1$ is a function
symbol in $\F{S_1}$, then the corresponding operator
$\hat{A_1}\in\mathcal{B(H}_1)_{\sa}$ will be `translated' to the
operator $\hat{A_1}\otimes\hat{1}_{\Hi_2}\in
\mathcal{B(H}_{1}\otimes\Hi_2)$. By assumption, this corresponds
to the function symbol $A_1\di1$ in $\F{S_{1}\di S_{2}}$.

\paragraph{Operator entanglement and translations.} We should be
cautious about what to expect from this translation when we
represent a physical quantity $A:\Si\map\R$ in $\F{S_1}$ by an
arrow between presheaves, since there are no canonical projections
\begin{equation}
\Hi_1\leftarrow\Hi_1\otimes\Hi_2 \rightarrow\Hi_2,
\end{equation}
and hence no canonical projections
\begin{equation}
\Sig^{\Hi_1}\leftarrow\Sig^{\Hi_1\otimes\Hi_2}\rightarrow
\Sig^{\Hi_2}
\end{equation}
from the spectral presheaf of the composite system to the spectral
presheaves of the components.\footnote{On the other hand, in the
classical case, there \emph{are} canonical projections
\begin{equation}
\Si_{\s,S_{1}}\leftarrow\Si_{\s,S_{1}\di S_{2}}
\rightarrow\Sigma_{\s,S_{2}}
\end{equation}
because the symplectic manifold $\Si_{\s,S_{1}\di S_{2}}$ that
represents the composite system is the cartesian product
$\Si_{\s,S_{1}\di S_{2}}=\Si_{\s,S_{1}}\times\Si_{\s,S_{2}}$,
which is a product in the categorial sense and hence comes with
canonical projections.}

This is the point where a form of \emph{entanglement} enters the
picture. The spectral presheaf $\Sig^{\Hi_1\otimes\Hi_2}$ is a
presheaf over the context category
$\mathcal{V}(\Hi_1\otimes\Hi_2)$ of $\Hi_1\otimes\Hi_2$. Clearly,
the context category $\V{1}$ can be embedded into
$\mathcal{V}(\Hi_1\otimes\Hi_2)$ by the mapping $V_1\mapsto
V_1\otimes\mathC\hat 1_{\Hi_2}$, and likewise $\V{2}$ can be
embedded into $\mathcal{V}(\Hi_1\otimes\Hi_2)$. But not every
$W\in\Ob{\mathcal{V}(\Hi_1\otimes\Hi_2)}$ is of the form
$V_1\otimes V_2$.

This comes from the fact that not all vectors in
$\Hi_1\otimes\Hi_2$ are of the form $\psi_1\otimes \psi_2$, hence
not all projections in $\mathcal{P}(\Hi_1\otimes\Hi_2)$ are of the
form $\P_{\psi_1}\otimes\P_{\psi_2}$, which in turn implies that
not all $W\in\mathcal{V}(\Hi_1\otimes\Hi_2)$ are of the form
$V_1\otimes V_2$. There are more contexts, or world-views,
available in $\mathcal{V}(\Hi_1\otimes\Hi_2)$ than those coming
from $\V{1}$ and $\V{2}$. We call this `operator entanglement'.

The topos representative of $\hat{A_1}$ is
$\dasB{A_1}:\Sig^{\Hi_1}\map \SR{}^{\;\Hi_1}$, and the
representative of $\hat{A_1}\otimes\hat{1}_{\Hi_2}$ is
$\breve\delta(A_1\di 1):
\Sig^{\Hi_1\otimes\Hi_2}\map\SR{}^{\;\Hi_1\otimes\Hi_2}$. At
subalgebras $W\in\Ob{\mathcal{V(H}_{1}\otimes\Hi_2)}$ which are
\emph{not} of the form $W=V_1\otimes V_2$ for any
$V_1\in\Ob{\mathcal{V(H}_{1})}$ and
$V_2\in\Ob{\mathcal{V(H}_{2})}$, the daseinised operator
$\delta(\hat{A_1}_{W}\otimes\hat{1}_{\Hi_2})\in W_{\sa}$ will not
be of the form
$\delta(\hat{A_1})_{V_1}\otimes\delta(\hat{A_1})_{V_2}$.\footnote{Currently,
it is even an open question if
$\delta(\hat{A_1}_{W}\otimes\hat{1}_{\Hi_2})=
\delta(\hat{A_1})_{V_1}\otimes\hat{1}_{\Hi_2}$ if $W=V_1\otimes
V_2$ for a non-trivial algebra $V_2$.} On the other hand, it is
easy to see that  $\delta(\hat{A_1}\otimes\hat{1}_{\Hi_2})_{W}=
\delta(\hat{A_1})_{V_1}\otimes\hat{1}_{\Hi_2}$ if
$W=V_1\otimes\mathC\hat 1_{\Hi_2}$.

Given a physical quantity $A_1$, represented by the arrow
$\dasB{A_1}:\Sig^{\Hi_1}\rightarrow\SR^{\Hi_1}$, we can (at best)
expect that the translation of this arrow into an arrow from
$\Sig^{\Hi_1\otimes\Hi_2}$ to $\SR^{\Hi_1\otimes\Hi_2}$ coincides
with the arrow $\dasB{A_1\di 1}$ on the `\emph{image}' of
$\Sig^{\Hi_1}$ in $\Sig^{\Hi_1\otimes\Hi_2}$. This image will be
constructed below using a certain geometric morphism. As one might
expect, the image of $\Sig^{\Hi_1}$ is a presheaf $\ps P$ on
$\mathcal{V}(\Hi_1\otimes\Hi_2)$ such that $\ps
P_{V_1\otimes\mathC\hat 1_{\Hi_2}}\simeq\Sig^{\Hi_1}_{V_1}$ for
all $V_1\in\V{1}$, \ie\ the presheaf $\ps P$ can be identified
with $\Sig^{\Hi_1}$ exactly on the image of $\V{1}$ in
$\mathcal{V}(\Hi_1\otimes\Hi_2)$ under the embedding $V_1\mapsto
V_1\otimes\mathC\hat 1_{\Hi_2}$. At these stages, the translation
of $\dasB{A_1}$ will coincide with $\dasB{A_1\di 1}$. At other
stages $W\in\mathcal{V}(\Hi_1\otimes\Hi_2)$, the translation
cannot be expected to be the same natural transformation as
$\dasB{A\di 1}$ in general.

\paragraph{A geometrical morphism and a possible translation.}
The most natural approach to a translation is the following. Let
$W\in\Ob{\mathcal{V(H}_{1}\otimes\Hi_2)}$, and define $V_{W}
\in\Ob{\V{1}}$ to be the \emph{largest} subalgebra of
$\mathcal{B(H}_{1})$ such that
$V_{W}\otimes\,\mathC\hat{1}_{\Hi_2}$ is a subalgebra of $W$.
Depending on $W$, $V_{W}$ may, or may not, be the trivial
subalgebra $\mathC\hat{1}_{\Hi_1}$. We note that if
$W^{\prime}\subseteq W$, then
\begin{equation}
V_{W^{\prime}}\subseteq V_{W}, \label{VW'<VW}
\end{equation}
but $W^{^{\prime}}\subset W$ only implies $V_{W^{\prime}}\subseteq
V_{W}$.

The trivial algebra $\mathC\hat{1}_{\Hi_1}$ is not an object in
the category $\V{1}$. This is why we introduce the `augmented
context category' $\V{1}_*$, whose objects are those of $\V{1}$
united with $\mathC\hat{1}_{\Hi_1}$, and with the obvious
morphisms ($\mathC\hat{1}_{\Hi_1}$ is a subalgebra of all
$V\in\V{1}$).

Then there is a functor $n:{\mathcal{V(H}_{1}\otimes\Hi_2})\map
\V{1}_*$, defined as follows. On objects,
\begin{eqnarray}
        n:\Ob{\mathcal{V(}\Hi_1\otimes\Hi_2)} &\map& \Ob{\V{1}_*}                  \nonumber\\
    W &\mapsto& V_{W},
\end{eqnarray}
and if $i_{W^{\prime}W}:W^\prime\map W$ is an arrow in
$\mathcal{V(}\Hi_1 \otimes\Hi_2)$, we define $n(i_{W^{\prime}W}):=
i_{V_{W^{\prime}}V_{W}}$ (an arrow in $\V{1}_*$); if
$V_{W^{\prime}}=V_{W}$, then $i_{V_{W^{\prime}}V_{W}}$ is the
identity arrow $\operatorname*{id}_{V_{W}}$.

Now let
\begin{equation}
\nu:\Set^{\mathcal{V(H}_{1}\otimes\Hi_2)^{\op}}\map\Set^{{(\V{1}}_*)^{\rm
op}}
\end{equation}
denote the geometric morphism induced by $\pi$. Then the
(left-exact) inverse-image functor
\begin{equation}
\nu^*:\Set^{{(\V{1}}_*)^{\rm
op}}\map\Set^{\mathcal{V(H}_{1}\otimes\Hi_2)^\op}
\end{equation}
acts on a presheaf $\ps{F}\in\Set^{{(\V{1}}_*)^{\rm op}}$ in the
following way. For all $W\in\Ob{\mathcal{V(H}_{1}\otimes\Hi_2)}$,
we have
\begin{equation}
(\nu^*\ps{F})_W=\ps{F}_{n^{\op}(W)}=\ps{F}_{V_{W}}
\end{equation}
and
\begin{equation}
(\nu^*\ps{F})(i_{W^{\prime}W})=
                \ps{F}(i_{V_{W^{\prime}}V_{W}})
\end{equation}
for all arrows $i_{W^{\prime}W}$ in the category
$\mathcal{V(H}_{1}\otimes \Hi_2)$.\footnote{We remark, although
will not prove here, that the inverse-image presheaf $\nu^*\ps{F}$
coincides with the direct image presheaf $\phi_*\ps{F}$ of
$\ps{F}$ constructed from the geometric morphism $\phi$ induced by
the functor
\begin{eqnarray}
\ka:\mathcal{V(H}_{1})&\map&\mathcal{V(H}_{1}\otimes\Hi_2)
                                                \nonumber\\
V  &  \mapsto& V\otimes\mathC\hat{1}_{\Hi_2}.
\end{eqnarray}
Of course, the inverse image presheaf $\beta^*\ps{F}$ is much
easier to construct.}

In particular, for all $W\in\mathcal{V(H}_{1}\otimes\Hi_2)$, we
have
\begin{eqnarray}
(\nu^*\Sig^{\Hi_1})_W&=&\Sig^{\Hi_1}_{V_{W}},\label{Defnu*Sig}\\
(\nu^*\SR{}^{\;\Hi_1})_W  & =&\SR{}^{\;\Hi_1}_{V_{W}}.
\end{eqnarray}
Since $V_W$ can be $\mathC\hat 1_{\Hi_1}$, we have to extend the
definition of the spectral presheaf $\Sig^{\Hi_1}$ and the
quantity-value presheaf $\SR^{\Hi_1}$ such that they become
presheaves over $\V{1}_*$ (and not just $\V{1}$). This can be done
in a straightforward way: the Gel'fand spectrum $\Sig_{\mathC\hat
1_{\Hi_1}}$ of $\mathC\hat 1_{\Hi_1}$ consists of the single
spectral element $\l_1$ such that $\l_1(\hat 1_{\Hi_1})=1$.
Moreover, $\mathC\hat 1_{\Hi_1}$ has no subalgebras, so the
order-reversing functions on this algebra correspond bijectively
to the real numbers $\mathR$.

Using these equations, we see that the arrow
$\dasB{A_1}:\Sig^{\Hi_1}\map\SR{}^{\;\Hi_1}$ that corresponds to
the self-adjoint operator $\hat{A_1}\in\mathcal{B(H}_{1})_{\sa}$
gives rise to the arrow
\begin{equation}
\nu^*(\dasB{A_1}):\nu^*\Sig^{\Hi_1}\map\nu^*\SR{}^{\;\Hi_1}.
                                        \label{dum1}
\end{equation}
In terms of our earlier notation, the functor
$\tau_\phi(p_1):\SetH{1}\map
\Set^{\mathcal{V(H}_{1}\otimes\Hi_2)^\op}$ is $\nu^*$, and the
arrow in \eq{dum1} is the arrow
$\tau_\phi(j)(A_{\phi,S}):\tau_\phi(j)(\Si_{\phi,S})\map
\tau_\phi(j)(\R_{\phi,S})$ in \eq{PBPhiA2} with $j:S_1\map S$
being replaced by  $p:S_1\di S_2\map S_1$, which is the arrow in
$\Sys$ whose translation representation we are trying to
construct.

The next arrow we need is the one denoted
$\beta_\phi(j):\tau_\phi(j)(\R_{\phi,S})\map\R_{\phi,S_1}$ in
\eq{PBPhiA2}. In the present case, we define
$\beta_\phi(p):\nu^*\SR{}^{\;\Hi_1}
\map\SR{}^{\;\Hi_1\otimes\Hi_2}$ as follows. Let
$\alpha\in(\nu^*\SR{}^{\;\Hi_1})_W\simeq\SR{}^{\;\Hi_1}_{V_W}$ be
an order-reversing real-valued function on $\downarrow\!\! V_W$.
Then we define an order-reversing function
$\beta_\phi(p)(\alpha)\in \SR{}^{\;\Hi_1\otimes\Hi_2}_W$ as
follows. For all $W^\prime\subseteq W$, let
\begin{equation}
        [\beta_\phi(p)(\alpha)](W^\prime):=\alpha(V_{W^\prime})
\end{equation}
which, by virtue of \eq{VW'<VW}, is an order-reversing function
and hence a member of $\SR{}^{\;\Hi_1\otimes\Hi_2}_W$.

We also need an arrow in
$\Set^{\mathcal{V(H}_{1}\otimes\Hi_2)^\op}$ from
$\Sig^{\Hi_1\otimes\Hi_2}$ to $\nu^*\Sig^{\Hi_1}$, where
$\nu^*\Sig^{\Hi_1}$ is defined in \eq{Defnu*Sig}. This is the
arrow denoted
$\phi(j):\Si_{\phi,S_1}\map\tau_\phi(j)(\Si_{\phi,S})$ in
\eq{PBPhiA2}.

The obvious choice is to restrict
$\l\in\Sig^{\Hi_{1}\otimes\Hi_2}_W$ to the subalgebra
$V_{W}\otimes\mathC \hat{1}_{\Hi_2}\subseteq W$, and to identify
$V_{W}\otimes\mathC \hat{1}_{\Hi_1}\simeq
V_{W}\otimes\hat{1}_{\Hi_1}\simeq V_W$ as von Neumann algebras,
which gives
$\Sig^{\Hi_1\otimes\Hi_2}_{V_{W}\otimes\mathC\hat{1}_{\Hi_2}}
\simeq\Sig^{\Hi_1}_{V_{W}}$. Let
\begin{eqnarray}
\phi(p)_W:\Sig^{\Hi_1\otimes\Hi_2}_W &\map&
                            \Sig^{\Hi_1}_{V_{W}}\nonumber\\
                                \l &  \mapsto&\l|_{V_{W}}
\end{eqnarray}
denote this arrow at stage\ $W$. Then
\begin{equation}
\beta_\phi(p)\circ\nu^*(\dasB{A_1})\circ\phi(p):
\Sig^{\Hi_1\otimes\Hi_2} \map\SR{}^{\;\Hi_1\otimes\Hi_2}
\end{equation}
is a natural transformation which is defined for all
$W\in\Ob{\mathcal{V(H}_{1}\otimes\Hi_2)}$ and all $\l\in W$ by
\begin{eqnarray}
                \left(\beta_\phi(p)\circ\nu^*(\dasB{A_1})\circ\phi(p)\right)_W(\l)
                &=& \nu^*(\dasB{A})(\l|_{V_{W}})\\
                &=& \{V^\prime\mapsto\l|_{V^{\prime}}(\dasto{V^{\prime}}{A})\mid
V^{\prime}\subseteq V_{W}\}.
\end{eqnarray}
This is clearly an order-reversing real-valued function on the set
$\downarrow\!\! W$ of subalgebras of $W$, \ie\ it is an element of
$\SR{}^{\;\Hi_1\otimes\Hi_2}_W$. We define
$\beta_\phi(p)\circ\nu^*(\dasB{A_1})\circ\phi(p)$ to be the
translation representation, $\phi(\L{p})(\dasB{A_1})$ of
$\dasB{A_1}$ for the composite system.

Note that, by construction, for each $W$, the arrow
$(\beta_\phi(p)\circ\nu^*(\dasB{A_1})\circ\phi(p))_W$ corresponds
to the self-adjoint operator
$\delta(\hat{A_1})_{V_{W}}\otimes\hat{1}_{\Hi_2}\in W_{\sa}$,
since
\begin{equation}
\l|_{V_{W}}(\delta(\hat{A_1})_{V_{W}})=\l(\delta(\hat{A_1})_{V_{W}
}\otimes\hat{1}_{\Hi_2})
\end{equation} for all $\l\in\Sig^{\Hi_1\otimes\Hi_2}_W$.

\paragraph{Comments on these results.}
This is about as far as we can get with the arrows associated with
the composite of two quantum systems.  The results above can be
summarised in the equation
\begin{equation}                    \label{Eq_QCompSysTranslation}
        \phi(\L{p})(\dasB{A_1})_W=
\breve\delta(A_1)_{V_{W}}\otimes\hat{1}_{\Hi_2}
\end{equation}
for all contexts $W\in\Ob{\mathcal{V(H}_{1}\otimes\Hi_2)}$. If
$W\in\Ob{\mathcal{V}(\Hi_1\otimes\Hi_2)}$ is of the form
$W=V_1\otimes\mathC\hat 1_{\Hi_2}$, \ie\ if $W$ is in the image of
the embedding of $\V{1}$ into $\mathcal{V}(\Hi_1\otimes\Hi_2)$,
then $V_W=V_1$ and the translation formula gives just what one
expects: the arrow $\dasB{A_1}$ is translated into the arrow
$\dasB{A_1\di 1}$ at these stages, since
$\delta(\hat{A_1}\otimes\hat{1}_{\Hi_2})_{V_1\otimes\mathC\hat
1_{\Hi_2}}=
\delta(\hat{A_1})_{V_1}\otimes\hat{1}_{\Hi_2}$.\footnote{To be
precise, both the translation $\phi(\L{p})(\dasB{A_1})_W$, given
by (\ref{Eq_QCompSysTranslation}), and $\dasB{A\di 1}_W$ are
mappings from $\Sig^{\Hi_1\otimes\Hi_2}_W$ to
$\SR^{\Hi_1\otimes\Hi_2}_W$. Each
$\l\in\Sig^{\Hi_1\otimes\Hi_2}_W$ is mapped to an order-reversing
function on $\downarrow\!\! W$. The mappings
$\phi(\L{p})(\dasB{A_1})_W$ and $\dasB{A\di 1}_W$ coincide at all
$W^{\prime}\in\downarrow\!\! W$ that are of the form
$W^{\prime}=V^{\prime}\otimes\mathC\hat 1_{\Hi_2}$.}

If $W\in\Ob{\mathcal{V}(\Hi_1\otimes\Hi_2)}$ is not of the form
$W=V_1\otimes\mathC\hat 1_{\Hi_2}$, then it is relatively easy to
show that
\begin{equation}
      \delta(\hat{A_{1}}\otimes\hat{1}_{\Hi_2})_{W}\neq
      \delta(\hat{A_1})_{V_{W}}\otimes \hat{1}_{\Hi_2}
\end{equation}
in general. Hence
\begin{equation}
    \phi(\L{p})(\dasB{A_1}) \neq  \breve\delta(A_1\di 1), \label{phiLpdAn=}
\end{equation}
whereas, intuitively, one might have expected equality. Thus the
`commutativity' condition \eq{ClassCom} is not satisfied.

In fact, there appears to be \emph{no} operator $\hat
B\in\mathcal{B(H}_{1}\otimes\Hi_2)$ such that
$\phi(\L{p})(\dasB{A_1})=\dasB{B}$. Thus the quantity,
$\beta_\phi(p)\circ\nu^*(\dasB{A_1})\circ\phi(p)$, that is our
conjectured pull-back, is  an arrow in
$\Hom{\Set^{\mathcal{V(H}_{1}}\otimes\Hi_2)^{\op}}
{\Sig^{\Hi_1\otimes\Hi_2}}{\SR{}^{\;\Hi_1\otimes\Hi_2}}$ that is
not of the form $A_{\phi,S_1\di S_2}$ for any physical quantity
$A\in\F{S_1\di S_2}$.

Our current understanding is that this translation is `as good as
possible': the arrow
$\dasB{A_1}:\Sig^{\Hi_1}\rightarrow\SR^{\Hi_1}$ is translated into
an arrow from $\Sig^{\Hi_1\otimes\Hi_2}$ to
$\SR^{\Hi_1\otimes\Hi_2}$ that coincides with $\dasB{A_1}$ on
those part of $\Sig^{\Hi_1\otimes\Hi_2}$ that can be identified
with $\Sig^{\Hi_1}$. But $\Sig^{\Hi_1\otimes\Hi_2}$ is much
larger, and it is not simply a product of $\Sig^{\Hi_1}$ and
$\Sig^{\Hi_2}$. The context category
$\mathcal{V}(\Hi_1\otimes\Hi_2)$ underlying
$\Sig^{\Hi_1\otimes\Hi_2}$ is much richer than a simple product of
$\V{1}$ and $\V{2}$. This is due to a kind of operator
entanglement. A translation can at best give a faithful picture of
an arrow, but it cannot possibly `know' about the more complicated
contextual structure of the larger category.

Clearly, both technical and interpretational  work remain to be
done.

\section{Conclusions}
\label{Sec:conclusions} In the previous three papers we have
developed the idea that, for a given theory-type, the theory of a
particular system, $S$, is to be constructed in the framework of a
certain, system-dependent, topos. The central idea is that a local
language, $\L{S}$, is attached to each system $S$, and that the
application of a given theory-type to $S$ is equivalent to finding
a representation, $\phi$, of $\L{S}$ in a topos $\tau_\phi(S)$.

Physical quantities are represented by arrows in the topos from
the state object $\Si_{\phi,S}$ to the quantity-value object
$\R_{\phi,S}$, and propositions are represented by sub-objects of
the state object. The idea of a `truth sub-object' of
$P\Si_{\phi,S}$ then leads to a neo-realist interpretation of
propositions in which each proposition is assigned a truth value
that is a global element of the sub-object classifier
$\Omega_{\tau_\phi(S)}$.  In general, neo-realist statements about
the world/system $S$ are to be expressed in the internal language
of the topos $\tau_\phi(S)$. Underlying this is the
intuitionistic, deductive logic provided by the local language
$\L{S}$.

Every classical system uses the same topos, namely the topos of
sets. However, in general, the topos will be system dependent,
which leads to the problem of  understanding how the topoi for a
class of systems behave under the action of taking a sub-system,
or combining a pair of systems to give a single composite system.
In the present paper, we have presented a set of axioms that
capture the general ideas we are trying to develop. Of course,
these axioms are not cast in stone, and are still partly
`experimental' in nature. However, we have shown that classical
physics exactly fits our suggested scheme, and that quantum
physics `almost' does: `almost' because of the issues concerning
the translation representation of the arrows associated with
compositions of systems that were discussed in Section
\ref{SubSec:TranslCompQT}.

\paragraph{Is there `un gros topos'?} It is clear that there are
many topics for future research, both in regard to the first three
papers, and  to the present one. As far as the present paper is
concerned, a question that is of particular interest is if there
is a \emph{single} topos within which all systems of  a given
theory-type can be discussed. For example, in the case of quantum
theory the relevant topoi are of the form $\SetH{}$, where $\Hi$
is a Hilbert space, and  the question is whether all such topoi
can be gathered together to form a single topos (what Grothendieck
termed `un gros topos') within which all quantum systems can be
discussed.

There are well-known examples of such constructions in the
mathematical literature. For example, the category, ${\rm Sh}(X)$,
of sheaves on a topological space $X$ is a topos, and there are
collections $\bf T$ of topological spaces which form a
Grothendieck site, so that the topos ${\rm Sh}(\bf{T})$ can be
constructed. A particular object  in ${\rm Sh}(\bf{T})$ will then
be a sheaf over $\bf T$ whose stalk over any object $X$ in $\bf T$
will be the topos  ${\rm Sh}(X)$.

For our purposes, the ideal situation would be if the various
categories of systems, $\Sys$, can be chosen in such a way that
${\cal M}(\Sys)$ is a site. Then the topos of sheaves, ${\rm
Sh}({\cal M}(\Sys)),$ over this site would provide a common topos
in which all systems of this theory type---\ie\ the objects of
$\Sys$---can be discussed. We do not know if this is possible, and
it is a natural subject for future study.

\paragraph{Some more speculative lines of future research.} At a
conceptual level, one motivating desire for the entire research
programme was to find a formalism that would always give some sort
of `realist' interpretation, even in the case of quantum theory
which is normally presented in an instrumentalist way. But this
particular example raises an interesting point because the
neo-realist interpretation takes place in the topos $\SetH{}$,
whereas the instrumentalist interpretation works in the familiar
topos $\Set$ of sets, and one might wonder how universal is the
use of a pair of topoi in this way.

Another, related, issue concerns the representation of the
$\PL{S}$-propositions of the form $\SAin\De$  discussed in paper
II. This serves as a bridge between the `external' world of  a
background spatial structure, and the internal world of the topos.
This link is not present with the $\L{S}$ language whose
propositions are  purely internal terms of type $\O$ of the form
`$A(\va{s})\in\va{\De}$', as discussed in paper I. In a topos
representation, $\phi$, of $\L{S}$, these become propositions of
the form `$A\in\Xi$', where $\Xi$\ is a sub-object of $\R_\phi$.

In general, if we have an example of our axioms working
neo-realistically in a topos $\tau$, one might wonder if there is
an `instrumentalist' interpretation of the same theory in a
different topos, $\tau_i$, say? Of course, the word
`instrumentalism' is used metaphorically here, and any serious
consideration of such a pair $(\tau,\tau_i)$ would require a lot
of very careful thought.

However, if a pair $(\tau,\tau_i)$ does exist, the question then
arises of whether  there is a \emph{categorial} way of linking the
neo-realist and instrumentalist interpretations: for example, via
a functor $I:\tau\map\tau_i$. If so, is this related to some
analogue of the daseinisation operation that produced the
representation of the $\PL{S}$-propositions, $\SAin\De$ in quantum
theory? Care is needed in discussing such issues since informal
set theory is  used as a meta-language in constructing a topos,
and one has to be careful not to confuse this with the existence,
or otherwise, of an `instrumentalist' interpretation of any given
representation.

If such a functor, $I:\tau\map\tau_i$, did exist then one could
speculate  on the possibility of finding an `interpolating chain'
of functors
\begin{equation}
        \tau\map\tau^1\map\tau^2\map
        \cdots\map\tau^n\map \tau_i
\end{equation}
which could be interpreted conceptually as corresponding to an
interpolation between the philosophical views of  realism and
instrumentalism!

Even more speculatively one might wonder if ``one person's realism
is another person's instrumentalism''. More precisely, given a
pair $(\tau,\tau_i)$ in the sense above, could there be cases in
which the topos $\tau$  carries a neo-realism interpretation of a
theory with respect to an instrumentalist interpretation  in
$\tau_i$, whilst being the carrier of an instrumentalist
interpretation with respect to the neo-realism of a `higher'
topos; and so on? For example, is there some theory whose
`instrumentalist manifestation` takes place in the topos
$\SetH{}$?

 On the other hand,  one might want to say that `instrumentalist'
interpretations always take place in the world of classical set
theory, so that  $\tau_i$ should always be chosen to be $\Set$. In
any event, it would be interesting to study the quantum case more
closely to see if there are any categorial relations between the
formulation in $\SetH{}$ and the instrumentalism interpretation in
$\Set$. It can be anticipated that the action of daseinisation
will play an important role here. We hope to say more about this
in a  later paper.

\paragraph{Implications for quantum gravity.}
A serious claim stemming from our work is that a successful theory
of quantum gravity should be constructed in some topos $\cal
U$---the `topos of the universe'---that is \emph{not} the topos of
sets. All entities of physical interest will be represented in
this topos, including models for space-time (if there are any at a
fundamental level in quantum gravity) and, if relevant, loops,
membranes etc. as well as incorporating the anticipated
generalisation of quantum theory.

Such a theory of quantum gravity will have a neo-realist
interpretation in the topos $\cal U$, and hence would be
particularly useful in the context of quantum cosmology. However,
in practice, physicists  divide the world up into smaller, more
easily handled, chunks, and each of them would correspond to what
earlier we have called a `system' and, correspondingly, would have
its own topos. Thus $\cal U$ would be something like the `gros
topos' of the theory, and would combine together the individual
`sub-systems' in a categorial way. Of course, it is most unlikely
that there is any preferred way of dividing the universe up into
bite-sized chunks, but this is  not problematic as the ensuing
relativism would be naturally incorporated into the idea of a
Grothendieck site.

\bigskip
\noindent {\bf Acknowledgements} This research was supported by
grant RFP1-06-04 from The Foundational Questions Institute
(fqxi.org).  AD gratefully acknowledges financial support from the
DAAD.

This work is also supported in part by the EC Marie Curie Research
and Training Network ``ENRAGE'' (European Network on Random
Geometry) MRTN-CT-2004-005616.

\end{document}